\documentclass[fleqn,usenatbib]{mnras}

\usepackage{newtxtext,newtxmath}
\usepackage[flushleft]{threeparttable}
\usepackage[T1]{fontenc}
\DeclareRobustCommand{\VAN}[3]{#2}
\let\VANthebibliography\thebibliography
\def\thebibliography{\DeclareRobustCommand{\VAN}[3]{##3}\VANthebibliography}

\usepackage{graphicx}	

\usepackage{bm}  
\usepackage[displaymath, mathlines]{lineno}
\let\oldequation\equation
\let\oldendequation\endequation
\renewenvironment{equation}{\linenomathNonumbers\oldequation}{\oldendequation\endlinenomath}
  
\title{Model-Independent Determination of $H_0$ and $\Omega_{K,0}$ using Time-Delay Galaxy Lenses and Gamma-Ray Bursts}

\author[S. S. Du et al.]{
Shen-Shi Du,$^{1,5}$\thanks{duss.physics@gmail.com}
Jun-Jie Wei,$^{2,3}$
Zhi-Qiang You,$^{4,5}$
Zu-Cheng Chen,$^{4,5}$
Zong-Hong Zhu,$^{1}$\thanks{zhuzh@whu.edu.cn}
En-Wei Liang$^{6}$
\\
$^{1}$School of Physics and Technology, Wuhan University, Wuhan, Hubei 430072, China\\
$^{2}$Purple Mountain Observatory, Chinese Academy of Sciences, Nanjing 210023, China\\
$^{3}$School of Astronomy and Space Sciences, University of Science and Technology of China, Hefei 230026, China\\
$^{4}$Department of Astronomy, Beijing Normal University, Beijing 100875, China\\
$^{5}$Advanced Institute of Natural Sciences, Beijing Normal University, Zhuhai 519087, China\\
$^{6}$Guangxi Key Laboratory for Relativistic Astrophysics, School of Physical Science and Technology, Guangxi University, Nanning 530004, China
}

\date{Accepted XXX. Received YYY; in original form ZZZ}

\pubyear{2022}

\begin{document}
\label{firstpage}
\pagerange{\pageref{firstpage}--\pageref{lastpage}}
\maketitle

\begin{abstract}
Combining the `time-delay distance' ($D_{\Delta t}$) measurements from galaxy lenses and other distance indicators provides model-independent determinations of the Hubble constant ($H_0$) and spatial curvature ($\Omega_{K,0}$), only based on the validity of the Friedmann-Lema\^itre-Robertson-Walker (FLRW) metric and geometrical optics. To take the full merit of combining $D_{\Delta t}$ measurements in constraining $H_0$, we use gamma-ray burst (GRB) distances to extend the redshift coverage of lensing systems much higher than that of Type Ia Supernovae (SNe Ia) and even higher than quasars, whilst the general cosmography with a curvature component is implemented for the GRB distance parametrizations. Combining Lensing+GRB yields $H_0=71.5^{+4.4}_{-3.0}$~km s$^{-1}$Mpc$^{-1}$ and $\Omega_{K,0} = -0.07^{+0.13}_{-0.06}$ (1$\sigma$). A flat-universe prior gives slightly an improved $H_0 = 70.9^{+4.2}_{-2.9}$~km s$^{-1}$Mpc$^{-1}$. When combining  Lensing+GRB+SN Ia, the error bar $\Delta H_0$ falls by 25\%, whereas $\Omega_{K,0}$ is not improved due to the degeneracy between SN Ia absolute magnitude, $M_B$, and $H_0$ along with the mismatch between the SN Ia and GRB Hubble diagrams at $z\gtrsim 1.4$. Future increment of GRB observations can help to moderately eliminate the $M_B-H_0$ degeneracy in SN Ia distances and ameliorate the restrictions on cosmographic parameters along with $\Omega_{K,0}$ when combining Lensing+SN Ia+GRB. We conclude that there is no evidence of significant deviation from a (an) flat (accelerating) universe and $H_0$ is currently determined at 3\% precision. The measurements show great potential to arbitrate the $H_0$ tension between the local distance ladder and cosmic microwave background measurements and provide a relevant consistency test of the FLRW metric.

\end{abstract}


\section{Introduction}\label{Sec:introduction}

The precise value of Hubble constant ($H_0$) has emerged as one of the most vital cracks in modern cosmology, with beyond $5\sigma$ tension between the local (late-universe) determination from Cepheid variable-calibrated Type Ia supernovae (SNe Ia; \citealt{Riess2019ApJ-876-85, Riess2021ApJ...908L...6R}) and the high-redshift (early-universe) measurement by \emph{Planck} cosmic microwave background (CMB;  \citealt{Planck2018resultsVI}) observations based on the $\Lambda$ Cold Dark Matter ($\Lambda$CDM) cosmological model. 
This issue has spurred intense debate of beyond-$\Lambda$CDM dark energy component or early-universe new physics as well as unaccounted-for systematic effects in either or both of the  measurements (e.g., \citealt{Riess2011ApJ-730-119, Marra2013PhRvL.110x1305M,Verde2013PDU-2-166, Riess2016ApJ-826-56, Bernal2016JCAP-10-019, Riess2019NatRP-2-10,Verde2019NatAs-3-891, Freedman2019ApJ...882...34F, Guido2021JCAP...05..072D,Di2021CQGra..38o3001D}). 
Meanwhile, various recent measurements of present-day spatial curvature ($\Omega_{K,0}$) have been shown to mutually disagree with each other, leading to another potential cosmic crisis \citep{Valentino2020NatAs...4..196D}. 
New, independent measurements of the $H_0$ and $\Omega_{K,0}$ are pursued to provide relevant clues to such discrepancies (see \citealt{Di2021CQGra..38o3001D} for a recent review). 

Time delay cosmography has become a mature field with the combined strengths of modern monitoring and lens modeling of quasar strong gravitational lensing, which provides a novel measurement of $H_0$ using the time delays between multiple gravitationally- and geometrically-induced images (\citealt{Refsdal1964MNRAS.128..307R, Blandford1992, Treu2016A&ARv, Bonvin2017MNRAS, Suyu2018SSRv..214...91S, Oguri2019RPPh...82l6901O, Wong2020MNRAS.tmp.1661W, Birrer2020A&A...643A.165B, Denzel2021MNRAS.501..784D}). 
The statistical combination of six measurements on different quasar time-delay lenses had reached a 2.4\% measurement of $H_0=73.3^{+1.7}_{-1.8}$~km s$^{-1}$Mpc$^{-1}$ in the flat $\Lambda$CDM model \citep{Wong2020MNRAS.tmp.1661W}; Considering the mass-sheet transformations, which leave the lensing observables unchanged, as the dominant source of residual uncertainty in $H_0$ estimation, \cite{Birrer2020A&A...643A.165B} obtained $74.5^{+5.6}_{-6.1}$~km s$^{-1}$Mpc$^{-1}$ inferring from the TDCOSMO sample with seven lenses. 
The measurements are not only independent of the early- and late-universe estimates, 
but are also proposed to control the systematic errors -- around 40 strong lensing systems can measure $H_0$ at a sub-percent level \citep{Jee2016JCAP,Shajib2019MNRAS.483.5649S}, thus raising the prospect of adjudicating the tension between CMB and the local distance ladder. 
Nevertheless, time-delay cosmography relies on the assumption of a specific background cosmology, and the inferred value of $H_0$ could be significantly different with using different cosmological models (see Table~6 in \citealt{Wong2020MNRAS.tmp.1661W}). Thus, this measurement is strongly model-dependent.

A more intriguing application of time-delay galaxy lenses (TDGLs) is combining the TDGLs and such reliable distance indicators from other probes to constrain both the $H_0$ and $\Omega_{K,0}$ via the distance sum rule (DSR; \citealt{Collett2019PhRvL.123w1101C}), merely assuming that the light propagates along null geodesics in the Friedmann-Lema\^itre-Robertson-Walker (FLRW) metric. 
This method only relies on the validity of the classical FLRW metric and geometrical optics, and infers $H_0$ and $\Omega_{K,0}$ with TDGLs providing the  ``time-delay distance" ($D_{\Delta t}$) and with the distance indicators extending the redshifts to the lens and source to cover that of the lensing systems well. Additionally, the measurement of $\Omega_{K,0}$ from two or more lens-source pairs provides a consistency test of the FLRW metric \citep{Rasanen2015PhRvL.115j1301R}. 
Such measurements has been implemented in combinations with, for instance SNe Ia \citep{Collett2019PhRvL.123w1101C}, multi-messenger gravitational-wave ``standard sirens" \citep{Liao2019PhRvD..99h3514L}, and ultra-compact structure in radio sources \citep{Qi2020arXiv201100713Q}. 
We also notice that \cite{Wei2020ApJ} proposed using quasars (QSOs) as distance indicators up to $z\sim5$ to match the distances of lensing systems that can be detectable by the upcoming {Rubin Observatory Legacy Survey of Space and Time (LSST; \citealt{ Oguri2010MNRAS.405.2579O,Ivezic2019ApJ...873..111I}), which is expected to take full advantage of the lensing catalog in hand. }
{While recent works were indicative of the $\sim4\sigma$ tension between the quasars and flat $\Lambda$CDM at high redshifts (e.g., $z\geq1.5$) and showed the vigilance of using quasars as standardizable candles to constrain cosmological parameters (e.g., \citealt{Risaliti2019NatAs...3..272R, Lusso2020A&A...642A.150L, Yang2020PhRvD.102l3532Y, Velten2020PhRvD.101d3502V, Khadka2021MNRAS.tmp..510K}). 
This prompts us to further explore robust high-redshift distance indicators to consolidate the independent measurements of $H_0$ and $\Omega_{K,0}$ in conjunction with TDGLs via the DSR. 
}

Gamma-ray bursts (GRBs) are the most luminous astrophysical phenomenon, which are expected to be emerged as early as death of the first-generation stars ($z\sim 20$). 
The central engine and radiation mechanisms as well as the jet composition of GRBs, which are responsible for the energetic $\gamma$-ray emission, remain inconclusive. Some aspects of the progenitor models have been well investigated in light of the fact that several long-duration GRBs (LGRBs) are observed in association with core-collapse supernova explosions (see e.g., \citealt{Zhang2018pgrb.book.....Z}). 
Despite these difficulties, the high luminosity of GRBs and the immunity of $\gamma-$rays from dust extinctions contribute to their high detectabilities at distances that are far beyond that SNe Ia (or even quasars) are available, thus GRBs are promising objects serving as cosmological probes to 
provide meaningful cosmological constraints over a wide redshift range \citep{Dai2004ApJ...612L.101D,Liang2005ApJ...633..611L, Ghirlanda2006NJPh, Amati2008MNRAS.391..577A,Amati2019MNRAS.486L..46A, WangFY2015NewAR..67....1W, Demianski2017A&A...598A.112D, Demianski2021MNRAS.tmp.1707D, Khadka2021arXiv210512692K}. 
One of the most important observational properties of LGRBs is the phenomenological relations between the observed spectral and intensity quantities (radiated $\gamma-$ray energy or luminosity) during the prompt phase, among which the Amati correlation  \citep{Amati2002A&A} between the $\nu F_{\nu}$ spectral peak energy, $E_{\rm p}$, and the bolometric isotropic-equivalent radiated energy, $E_{\rm iso}$, has been extensively explored as the distance indicators of LGRBs (e.g., \citealt{Amati2008MNRAS.391..577A, Amati2019MNRAS.486L..46A,  Demianski2017A&A...598A.112D, Demianski2021MNRAS.tmp.1707D}). 
The main obstacles that preclude the cosmological use of GRBs come from the potential selection effects, the ``circularity problem" in calibrating the GRB luminosity correlations, and their possible redshift evolutionary effects (see a recent review by \citealt{WangFY2015NewAR..67....1W}). 
Several works have investigated the observational selection effects and show a general conclusion that such effects, even though still debated, might be minor (e.g., \citealt{Amati2006MNRAS.372..233A, Butler2009ApJ...694...76B, Nava2012MNRAS.421.1256N, Demianski2017A&A...598A.112D}). 
The circularity has been effectively overcome with several reliable calibration procedures \citep{WangFY2015NewAR..67....1W, Amati2019MNRAS.486L..46A}. 
We will elaborate on the impact of possible redshift evolutionary effects on the Amati correlation when modelling the GRB distance indicators.

In this work, we propose to combine the $D_{\Delta t}$ measurements from TDGLs and distances determined with the GRB Hubble diagram to restrict $H_0$ and $\Omega_{K,0}$ via the DSR. 
The general cosmography with an arbitrary spatial curvature component is implemented for the GRB distance parametrizations, with which the bounds of $H_0$ and $\Omega_{K,0}$ are contributed by both the GRB distances and $D_{\Delta t}$. 
To alleviate the circularity, we simultaneously fit the GRB distance indicators and the parameters in theoretical distances in an effective Bayesian framework, and $H_0$ and $\Omega_{K,0}$ can be obtained by marginalizing over the remaining parameters. 
The rest of this paper is arranged as follows. 
In Section~\ref{Sec:sec2}, we depict our method to determine the $H_0$ and $\Omega_{K,0}$ with the combined observations of TDGLs and the popular high-redshift probes, including the most recently released SNe Ia, GRBs, and QSOs, which we denote as standardizable-candle (SC) objects hereafter. 
The observational data and likelihood constructions are detailed in Section~\ref{Sec:sec3}. 
We present the results and discussions in Section~\ref{Sec:sec4}. 
Summary is made in Section~\ref{Sec:sec5}. 

\section{Methodology}\label{Sec:sec2}
\subsection{Constraining Hubble constant and spatial curvature through distance sum rule}

In what follows, we present the method for simultaneously estimating the $H_0$ and $\Omega_{K,0}$ with only the homogeneity and isotropy ansatz in cosmology, stating that the space-time is described by the FLRW metric, 
\begin{equation}
\label{Eq:metric}
{\rm d}s^2 = -c^2 t^2+ a^2(t)\left[\frac{{\rm d}r^2}{1-\kappa r^2}+ r^2d\Omega^2\right],
\end{equation}
where $c$ is the speed of light, $a(t)=(1+z)^{-1}$ is the scale factor, and the dimensionless curvature $\kappa$=1, 0, and $-1$ corresponds to a spatially closed, flat, and open universe, respectively. 
The cosmic expansion rate is defined as $H(z)\equiv{\dot a}(t)/a(t)$ with its present value is $H_0$ (hereafter, the subscript ``0" represents the present value), which acts as a normalization of the Hubble diagram. 
Let $D_{\rm A}(z_s, z_l)$ be the angular diameter distance of 
a source emerging at redshift $z_s$ as lensed at $z_l$, 
where subscripts ``$s$" and ``$l$" correspond to source and lens, respectively.
Assuming that the geometrical optics holds, 
the dimensionless co-moving distance, $d(z_s, z_l) \equiv (1+z_s)H_0D_{\rm A}(z_s, z_l)/c$, 
is given by
\begin{equation}\label{Eq:dzszl}
\begin{split}
d(z_s, z_l) =\frac{1}{\sqrt{|\Omega_{K,0}|}}{\rm sinx}\left({\sqrt{|\Omega_{K,0}|}}\int^{z_s}_{z_l}\frac{d\Tilde{z}}{E(\Tilde{z})}\right),
\end{split}
\end{equation}
where $E(z)\equiv H(z)/H_0$, the curvature density parameter $ \Omega_{K,0}\equiv -\kappa c^2/a_0^2H_0^2$, 
and sinx is sinh (sin) in the case of $\Omega_{K,0}>0$ ($\Omega_{K,0}<0$) and, for $\Omega_{K,0}=0$ this equation simplifies as the integral. 
{With the denotation of $d(z)\equiv d(0,z)$, Equation~(\ref{Eq:dzszl}) links $d_{ls}\equiv d(z_s, z_l)$,  $d_{s}\equiv d(z_s)$, and  $d_{l}\equiv d(z_l)$ in the FLRW metric via the distance sum rule \citep{Rasanen2015PhRvL.115j1301R}, }
\begin{equation}\label{Eq:dls}
d_{ls}=\epsilon_1 d_s \sqrt{1+\Omega_{K,0}d^2_l} - \epsilon_2 d_l \sqrt{1+\Omega_{K,0}d^2_s},
\end{equation}
where $\epsilon_i = \pm1$ with the positive case is required for the one-to-one correspondence between $z$ and $t$, with $d'(0)>0$.  
We assume the case of $\epsilon_i=1$, then Equation~(\ref{Eq:dls}) can be rewritten as
\begin{equation}\label{Eq:combination}
\frac{d_{ls}}{d_sd_l}=\sqrt{1/d_l^2+\Omega_{K,0}}- \sqrt{1/d_s^2+\Omega_{K,0}}.
\end{equation}

Strong lensing systems benefit from their very simple underlying physics 
and purely gravitational origin compared to the other cosmological probes. 
The measurement of time delay $\Delta t_{i,j}$ between lensed images $i$ and $j$ at the coordinates $\bm \theta_{i} $ and $\bm \theta_{j} $ and a concrete lens model provide the measurement of time-delay distance $D_{\Delta t}$ through  \citep{Refsdal1964MNRAS.128..307R, Schneider1992grle.book.....S}

\begin{equation}\label{Eq:timedelay}
\Delta t_{i,j}= \frac{D_{\Delta t}}{c}\Delta \psi_{i,j}, 
\end{equation}
where $\Delta \psi_{i,j}= [(\bm \theta_i-\bm \beta_{s})^2/2-\psi(\bm \theta_i)-(\bm \theta_j-\bm \beta_{s})^2/2+\psi(\bm \theta_{j})]$ represents the ``Fermat potential" difference between lensed images with the two-dimensional lensing potential $\psi$ and source position $\bm{\beta_{s}}$. 
The time-delay distance is the combination of the three angular diameter distances:
\begin{equation}\label{Eq:lensdistance}
 D_{\Delta t} = (1+z_l)\frac{D_{A,l} D_{A,s}}{D_{A,ls}}=\frac{c}{H_0}\frac{d_ld_s}{d_{ls}}.
 \end{equation} 
Therefore, $H_0$ and $\Omega_{K,0}$ can be directly determined by comparing Equation~(\ref{Eq:combination}) with the time-delay distance ratio $d_{ls}/{d_sd_l}$ extracted from lensing time delay data through Equations~(\ref{Eq:timedelay}) and (\ref{Eq:lensdistance}). 

To reduce the model dependence and take the full advantage of the whole $D_{\Delta t}$ measurements, we make use of GRBs as distance indicators to derive $d_l$ and $d_s$ out to very high redshifts in Equation~(\ref{Eq:combination}) via ${{d(z) =H_0d_{\rm L}(z)[c(1+z)]^{-1}}}$, which holds in any space-time for any gravity theory. Due to the lack of low-$z$ GRBs, we also combine the well-calibrated Pantheon SNe Ia (see Section~\ref{Sec:SC}) as references to optimize the constraints on $d_l$ and $d_s$.
In principle, one needs to select a pair of SC events that match the redshifts of the lens and source in a lensing system. 
However, it is difficult for this to be fulfilled with all the discrete data, and there are always differences in the lensing redshifts from the nearest SC objects. 
This can be resolved by smoothing the evolution of luminosity distances from all discrete events through a model-independent way, for instance, making use of a polynomial \citep{Rasanen2015PhRvL.115j1301R} or non-parametric approach like the Gaussian process \citep{Shafieloo2012PhRvD}. 
It is worth stressing that, the polynomial $d(z)=\sum_{i=1}^{n}c_i z^i$ conditional on $d(0)=0$ and $d'(0)=1$ adopted in previous works (e.g., \citealt{Collett2019PhRvL.123w1101C, Wei2020ApJ}) as the parametrization of SC distances might suffer from the divergence and convergence problems when fitting to the high-redshift data sets, and the finite truncated polynomials could result in some systematic errors. General cosmography provides an ideal tool to explore the dynamical evolution of cosmic expansion according to the cosmological principle and without specifying a cosmological model, intrinsically owning the idea of fitting data using the Taylor series\footnote{Analogous approaches have been proposed to parameterize cosmic distances, such as Pad\'e series \citep{Gruber2014PhRvD..89j3506G} and Chebyshev polynomials \citep{Capozziello2018MNRAS.476.3924C}. However, it is very difficult to determine the approximation order due to the dearth of data points and the qualities of cosmological measurements.} expanding around $z\simeq0$ with the scale factor derivatives up to a certain order \citep{Visser2004CQGra..21.2603V, Capozziello2013Galax...1..216C, Capozziello2019IJMPD..2830016C, Aviles2014PhRvD..90d3531A}. 

To avoid the convergence and divergence problem mentioned above, we adopt the Taylor series expansions, with the scale factor derivative up to the fourth order, as a function of the y-redshift \citep{Cattoen2008PhRvD..78f3501C}, $y=z/(1+z)$, for the SC distance parametrizations (see \citealt{Capozziello2013Galax...1..216C} for more details). 
In the Appendix, we define the leading-order cosmographic parameters to be constrained, including the deceleration $q_0$, jerk $j_0$, snap $s_0$ parameters, etc., and different distances determined by Taylor series are presented (see \citealt{Capozziello2013Galax...1..216C} for more detailed descriptions about the evolutionary status of the universe corresponding to these parameters). Note that the SC distances encode both the $H_0$ and $\Omega_{K,0}$ in the cosmography formulae. Thus our method can adequately explore how constraining the $H_0$ and $\Omega_{K,0}$ are in the DSR approach. 

\subsection{Bayesian framework}
We infer the free parameters within the Bayesian framework (the methods described below are referred to \citealt{Trotta2017arXiv170101467T}). 
Given the observational data, ${\bm D}$, and prior knowledge in modelling the SC distance indicators (see Section~\ref{Sec:SC}) and cosmographic models (i.e., the order of expansions), the posterior probability distribution of the free parameters, ${\bm \theta}$, is given by
\begin{equation}\label{Eq:Bayes}
p({\bm \theta | {\bm D}})=\frac{{\mathcal L}({\bm \theta}; {\bm D}) \pi ({\bm \theta})}{\int {{\mathcal L}({\bm \theta}; {\bm D}) \pi({\bm \theta})} d{\bm \theta}}.
\end{equation}
The term ${\mathcal L}({\bm \theta}; {\bm D})=p({\bm D}|{\bm \theta})$ is the \emph{likelihood} of observational data conditional on the knowledge of such free parameters of hypothetic models, which permits to combine different data sets with the summation of log-likelihoods for the parameter inference. 
For the \emph{prior} $\pi ({\bm \theta})$, we use uniform distributions by taking large and physically acceptable ranges for each free parameter unless specified otherwise, as shown in Table~\ref{MyTabA}.
The right-hand-side integral in Equation~(\ref{Eq:Bayes}) is the \emph{evidence}.

The marginalized posterior probability distribution of each parameter is obtained with implementing the dynamical nested sampling code {\tt dynesty}  \citep{Speagle2020MNRAS.493.3132S}\footnote{\url{https://github.com/joshspeagle/dynesty}}. 
Dynamical nested sampling has desirable merits over the Markov chain Monte Carlo algorithms with sophisticated treatment of multimodal solutions, more flexible access to independent samples, and in particular, focusing on effectively and simultaneously estimating the posterior and evidence while sampling from complicated, multimodal distributions. Besides, dynamic nested sampling allows samples to be targeted adaptively to better sample specific areas of the posterior during the fit. In this work, for each Bayesian parameter, we report the median of the marginalized posterior probability distribution along with the $1\sigma$ error bar taken as half of the 16th$-$84th quantile range.

\section{Data and likelihood constructions}\label{Sec:sec3}

\subsection{Strong Lensing Time-Delay Distances}\label{Sec:Lens}

We compile the recent six strongly lensed quasars 
with time-delay measurements from H0LiCOW collaboration \citep{Wong2020MNRAS.tmp.1661W}, { including B1608+656 \citep{Suyu2010ApJ...711..201S,Jee2019Sci...365.1134J}, RXJ1131-1231 \citep{Suyu2014ApJ...788L..35S, Chen2019MNRAS.490.1743C}, HE0435-1223 \citep{Wong2017MNRAS.465.4895W,Chen2019MNRAS.490.1743C}, WFI2033-4723 \citep{Rusu2020MNRAS.498.1440R}, J1206+4332 \citep{Birrer2019MNRAS.484.4726B}, and PG1115+080 \citep{Chen2019MNRAS.490.1743C}.} 
All lenses except B1608+656 are blindly analyzed with respect to the cosmological parameters. 
The posterior distributions of time-delay distance of these galaxy lenses are publicly released at the H0LiCOW website\footnote{\url{http://www.h0licow.org}}. 
{ For B1608+656, the time-delay distance likelihood function was approximated with a skewed log-normal distribution}:
\begin{equation}
{\mathcal L}_{D_{\Delta \tau}}=\frac{1}{\sqrt{2\pi}(x-\lambda_D)\sigma_D}\exp\left[{-\frac{(\ln (x-\lambda_D)-\mu_D)^2}{2\sigma^2_D}}\right], 
\end{equation}
with the parameters $\mu_{\rm D}=7.0531$, $\sigma_{\rm D}=0.22824$, and $\lambda_{\rm D}=4000.0$, where $x=D_{\Delta t}$ (in units of Mpc). 
The time delay posterior distributions of the remaining five lensed quasars were released with the samples of Monte-Carlo Markov Chains. 
We also include DES-J0408-5354 from STRIDES collaboration \citep{Shajib2020MNRAS.494.6072S}, 
which sets the current most precise measurement of $H_0$ from a single time-delay lens. 
For this event, we use the time-delay distance posterior distribution derived by \cite{Shajib2020MNRAS.494.6072S}. 
Besides, SDSSJ0946+1006 \citep{Gavazzi2008ApJ...677.1046G} is a double-source-plane strong lensing with the presence of two sources ($z_{s1}=0.609$, $z_{s2}=2.3$) that lensed by the same foreground galaxy ($z_l=0.222$), 
offering an accurate constraint on the cosmological scaling factor $\xi=(d_{ls1}d_{s2}/d_{s1}d_{ls2})=(d_{ls1}/d_ld_{ls1})\cdot (d_ld_{l,s2}/d_{l,s2})$, which is sensitive to $\Omega_{K,0}$ and with no dependence on $H_0$.  
In SDSSJ0946+1006, the constrain result is $\xi^{-1}=1.404\pm 0.016$ \citep{Collett2014MNRAS.443..969C}. 
Thus, we can well approximate the likelihood function ${\mathcal L}_{\xi^{-1}}$ 
using a Gaussian function with the mean value and dispersion $\mu_{\xi^{-1}}\pm \sigma_{\xi^{-1}}=1.404\pm 0.016$, i.e.,
\begin{equation}
{\mathcal L}_{\xi^{-1}}=\frac{1}{\sqrt{2\pi}\sigma_{\xi^{-1}}}\exp\left\{-\frac{[\xi^{-1}-\frac{T(z_l)-T(z_{s2})}{T(z_l)-T(z_{s1})}]}{2\sigma^2_{\xi^{-1}}}\right\},
\end{equation}
with $T(z) = \sqrt{1/d^2(z) + \Omega_{K,0}}$.

\begin{table*} 
\small
\begin{center}
\setlength{\tabcolsep}{1em}
\caption{Priors used in the nested sampling and each free parameter's descriptions.  $\mathcal{U}$ represents a uniform probability distribution.}
\label{MyTabA}
    \begin{tabular}{lll}
    \hline\hline
    Parameter & Prior & Description of Parameter \\
    \hline
		$M_{\rm B}$~[mag] & $\mathcal U$(-25,-15) & B-band absolute peak magnitude of SNe Ia. \\
		$A$ & $\mathcal U$(0,100) &  Intercept of the Amati correlation of LGRBs in logarithm.\\
		$B$ & $\mathcal U$(0,5) & Slope of Amati correlation of LGRBs in logarithm.\\
		$\delta_{\rm grb}$ & $\mathcal U$(0,2) & Intrinsic dispersion of Amati correlation of LGRBs in logarithm.\\
		$\beta'$ & $\mathcal U$(-5,5) & Parameter related to the intercept ($\beta$) of quasar's $L_{\rm X}-L_{\rm UV}$ relation in logarithm.\\
		$\gamma$ & $\mathcal U$(0,2) & Slope of quasar's $L_{\rm X}-L_{\rm UV}$ relation in logarithm.\\
		$\delta_{\rm qso}$ & $\mathcal U$(0, 2) & Intrinsic scatter of quasar's $L_{\rm X}-L_{\rm UV}$ relation in logarithm.\\
		$H_0$~[km$\cdot$s$^{-1}\cdot$Mpc$^{-1}$] & $\mathcal U$(40, 90) & Present cosmic expansion rate.\\
		$\Omega_{K,0}$ & $\mathcal U$(-1, 1) & Present energy density of spatial curvature component.\\
		$q_0$ & $\mathcal U$(-2,0) & Cosmographic deceleration parameter. \\
		$j_0$ & $\mathcal U$(-10,10) & Cosmographic jerk parameter. \\
		$s_0$ & $\mathcal U$(-150,150) & Cosmographic snap parameter.\\
		\hline
    \end{tabular}
\end{center}
\end{table*}

\begin{table*}
\small
\begin{center}
\setlength{\tabcolsep}{1em}
    \centering
    \caption{The bounds obtained with different independent data combinations at 1$\sigma$ credibility.}
    \label{MyTabB}
    \begin{tabular}{lccccccccc}
        \hline
		Data combined & $H_0^{*}$ & $\Omega_{K,0}$ & $q_0$ & $j_0$ & $s_0$ & $M_B$ & $A$/$\gamma$ & $B$/$\beta'$ & $\delta_{\rm grb}$/$\delta_{\rm qso}$ \\ 
		\hline
		Lensing+SN Ia & $75.8^{+3.7}_{-2.5}$ & $0.12^{+0.16}_{-0.14}$ & $-0.64^{+0.24}_{-0.25}$ & $2.6^{+3.9}_{-4.1}$ & $1^{+48}_{-27}$ & $-19.17^{+0.09}_{-0.08}$ & &  &  \\ 
		Lensing+SN Ia+SH0ES & $74.2^{+0.5}_{-1.4}$ & $0.04^{+0.10}_{-0.10}$ & $-0.56^{+0.25}_{-0.22}$ & $1.0^{+3.8}_{-3.6}$ & $-6^{+39}_{-25}$ & $-19.24^{+0.04}_{-0.02}$ \\ 
		Lensing+SN Ia+$H_0^{\star}$ & $76.6^{+2.5}_{-2.7}$ & $0.11^{+0.16}_{-0.11}$ & $-0.69^{+0.29}_{-0.19}$ & $2.7^{+3.7}_{-3.9}$ & $6^{+42}_{-31}$ & $-19.14^{+0.05}_{-0.09}$ \\ 
		Lensing+SN Ia & $73.8^{+2.3}_{-1.8}$ & flat & $-0.61^{+0.23}_{-0.29}$ & $2.1^{+3.6}_{-4.7}$ & $-22^{+61}_{-12}$ & $-19.24^{+0.06}_{-0.05}$ &  & &  \\
		Lensing+QSO & $77.6^{+2.8}_{-3.3}$ & $-0.21^{+0.41}_{-0.49}$ & $-0.41^{+0.28}_{-0.38}$ & $6.3^{+3.6}_{-2.7}$ & $34^{+41}_{-35}$ &  & $0.58^{+0.01}_{-0.01}$ & $-1.63^{+0.04}_{-0.02}$ & $0.22^{+0.01}_{-0.01}$ \\ 
		Lensing+QSO+$H_0^\star$ & $77.0^{+2.6}_{-2.6}$ & $-0.23^{+0.38}_{-0.50}$ & $-0.37^{+0.26}_{-0.35}$ & $7.6^{+1.8}_{-4.5}$ & $34^{+34}_{-40}$ &  &  $ 0.58^{+0.01}_{-0.01}$ & $-1.63^{+0.04}_{-0.02}$ & $0.22^{+0.00}_{-0.00}$ \\
		Lensing+QSO & $77.9^{+2.9}_{-2.5}$ & flat & $-0.45^{+0.34}_{-0.21}$ & $7.2^{+2.3}_{-3.7}$ & $18^{+48}_{-20}$ &  & $0.58^{+0.01}_{-0.01}$ & $-1.62^{+0.03}_{-0.03}$ &$0.22^{+0.01}_{-0.01}$  \\ 
		Lensing+GRB & $71.5^{+4.4}_{-3.0}$ & $-0.07^{+0.13}_{-0.06}$ & $-0.39^{+0.33}_{-0.67}$ & $-1.4^{+5.2}_{-4.9}$ & $96^{+44}_{-97}$ &  & $49.19^{+0.17}_{-0.18}$ & $1.47^{+0.06}_{-0.08}$ & $0.37^{+0.02}_{-0.02}$ \\ 
		Lensing+GRB+$H_0^\star$ & $72.9^{+3.1}_{-3.0}$ & $-0.02^{+0.17}_{-0.07}$ & $-0.69^{+0.43}_{-0.56}$ & $0.4^{+4.6}_{-5.1}$ & $56^{+59}_{-76}$ & &  $49.20^{+0.17}_{-0.17}$ & $1.46^{+0.07}_{-0.07}$ & $0.37^{+0.02}_{-0.02}$ \\ 
		Lensing+GRB & $70.9^{+4.2}_{-2.9}$ & flat & $-0.43^{+0.33}_{-0.53}$ & $-0.4^{+3.8}_{-5.2}$ & $98^{+46}_{-78}$ & & $49.21^{+0.15}_{-0.20}$ & $1.47^{+0.06}_{-0.08}$ & $0.37^{+0.03}_{-0.02}$ \\ 
		Lensing+GRB$^\dag$ & $70.4^{+3.5}_{-3.3}$ & $-0.07^{+0.03}_{-0.02}$ & $-0.43^{+0.32}_{-0.62}$ & $-3.3^{+3.4}_{-4.6}$ & $112^{+37}_{-62}$ &  & $49.31^{+0.10}_{-0.13}$ & $1.47^{+0.04}_{-0.04}$ & $0.27^{+0.01}_{-0.02}$ \\
		Lensing+GRB$^\dag$+$H_0^\star$ & $71.4^{+3.2}_{-2.8}$ & $-0.07^{+0.03}_{-0.02}$ & $-0.75^{+0.51}_{-0.44}$ & $-3.8^{+4.6}_{-3.9}$ & $112^{+37}_{-72}$ & & $49.31^{+0.11}_{-0.13}$ & $1.47^{+0.03}_{-0.05}$ & $0.27^{+0.01}_{-0.02}$ \\
		Lensing+GRB$^\dag$ & $69.9^{+3.0}_{-2.5}$ & flat & $-0.28^{+0.28}_{-0.32}$ & $-2.3^{+3.1}_{-3.2}$ & $133^{+16}_{-50}$ &  & $49.23^{+0.12}_{-0.09}$ & $1.47^{+0.04}_{-0.04}$ & $0.27^{+0.02}_{-0.02}$ \\ 
		Lensing+SN Ia+QSO & $78.0^{+2.9}_{-3.0}$ & $0.14^{+0.16}_{-0.15}$ & $-0.94^{+0.15}_{-0.11}$ & $8.9^{+1.0}_{-2.7}$ & $96^{+12}_{-39}$ & $-19.14^{+0.09}_{-0.07}$ & $0.60^{+0.01}_{-0.01}$ & $-1.56^{+0.01}_{-0.02}$ & $0.22^{+0.00}_{-0.00}$ \\
		Lensing+SN Ia+QSO+SH0ES & $74.1^{+0.8}_{-1.0}$ & $0.01^{+0.10}_{-0.10}$ & $-0.90^{+0.20}_{-0.12}$ & $7.8^{+1.8}_{-3.6}$ & $61^{+24}_{-34} $ & $-19.24^{+0.03}_{-0.03}$ & $0.60^{+0.00}_{-0.01}$ & $-1.54^{+0.01}_{-0.01}$ & $0.22^{+0.00}_{-0.00}$ \\
		Lensing+SN Ia+QSO+$H_0^{\star}$ & $77.0^{+2.7}_{-2.4}$ & $0.11^{+0.15}_{-0.14}$ & $-0.91^{+0.13}_{-0.14}$ & $8.5^{+1.4}_{-2.5}$ & $86^{+22}_{-31}$ & $-19.36^{+0.01}_{-0.02}$ & $0.60^{+0.00}_{-0.01}$ & $-1.52^{+0.01}_{-0.01}$ & $0.23^{+0.00}_{-0.00}$ \\
		Lensing+SN Ia+QSO & $75.4^{+2.3}_{-1.8}$ & flat & $-0.98^{+0.18}_{-0.11}$ & $9.1^{+0.9}_{-3.1}$ & $99^{+10}_{-47}$ & $-19.20^{+0.05}_{-0.06}$ & $ 0.59^{+0.01}_{-0.01}$ & $-1.59^{+0.01}_{-0.01}$ & $0.22^{+0.00}_{-0.00}$ \\
		Lensing+SN Ia+GRB & $75.6^{+3.3}_{-2.9}$ & $0.10^{+0.18}_{-0.12}$ & $-0.50^{+0.23}_{-0.24}$ & $0.7^{+3.7}_{-3.5}$ & $-10^{+40}_{-22}$ & $-19.18^{+0.09}_{-0.08}$ & $49.15^{+0.15}_{-0.20}$ & $1.38^{+0.07}_{-0.07}$ & $0.38^{+0.02}_{-0.02}$ \\ 
		Lensing+SN Ia+GRB+SH0ES & $74.0^{+0.7}_{-1.1}$ & $0.04^{+0.11}_{-0.08}$ & $-0.43^{+0.19}_{-0.23}$ & $-0.5^{+3.3}_{-2.9}$ & $-9^{+26}_{-24}$ & $-19.23^{+0.02}_{-0.03}$ & $49.15^{+0.16}_{-0.18}$ & $1.39^{+0.06}_{-0.07}$ & $0.38^{+0.02}_{-0.02}$ \\
		Lensing+SN Ia+GRB+$H_0^\star$ & $75.3^{+3.0}_{-2.5}$ & $0.10^{+0.16}_{-0.12}$ & $-0.52^{+0.28}_{-0.17}$ & $-1.2^{+4.8}_{-2.0}$ & $-17^{+38}_{-17}$ & $-19.19^{+0.08}_{-0.07}$ & $49.12^{+0.17}_{-0.17}$ & $1.40^{+0.05}_{-0.08}$ & $0.38^{+0.02}_{-0.02}$ \\ 
		Lensing+SN Ia+GRB & $73.3^{+2.2}_{-1.8}$ & flat & $-0.46^{+0.24}_{-0.26}$ & $-0.8^{+3.8}_{-3.7}$ & $-23^{+38}_{-18}$ & $-19.23^{+0.04}_{-0.06}$ & $49.17^{+0.15}_{-0.20}$ & $1.38^{+0.08}_{-0.06}$ & $0.38^{+0.02}_{-0.03}$ \\ 
		Lensing+SN Ia+GRB$^\dag$ & $74.6^{+2.5}_{-2.9}$ & $0.07^{+0.14}_{-0.12}$ & $-0.23^{+0.16}_{-0.17}$ & $-4.4^{+2.7}_{-2.0}$ & $-31^{+16}_{-15}$ & $-19.19^{+0.05}_{-0.10}$ & $48.98^{+0.11}_{-0.11}$ & $1.45^{+0.05}_{-0.03}$ & $0.29^{+0.02}_{-0.01}$ \\
		Lensing+SN Ia+GRB$^\dag$+SH0ES & $73.5^{+1.1}_{-0.8}$ & $0.04^{+0.1}_{-0.08}$ & $-0.19^{+0.12}_{-0.18}$ & $-4.6^{+2.6}_{-1.7}$ & $-30^{+14}_{-16}$ & $-19.23^{+0.02}_{-0.04}$ & $49.00^{+0.10}_{-0.12}$ & $1.47^{+0.04}_{-0.04}$ & $0.29^{+0.02}_{-0.01}$ \\ 
		Lensing+SN Ia+GRB$^\dag$+$H_0^{\star}$ & $74.8^{+2.3}_{-2.8}$ & $0.09^{+0.12}_{-0.14}$ & $-0.17^{+0.09}_{-0.22}$ & $-4.1^{+2.3}_{-2.1}$ & $-32^{+16}_{-14}$ & $-19.20^{+0.07}_{-0.08}$ & $48.98^{+0.11}_{-0.11}$ & $1.46^{+0.04}_{-0.04}$ & $0.29^{+0.02}_{-0.02}$ \\
		Lensing+SN Ia+GRB$^\dag$ & $73.2^{+1.6}_{-1.8}$ & flat & $-0.19^{+0.13}_{-0.20}$ & $-4.4^{+2.0}_{-2.4}$ & $-35^{+14}_{-14}$ & $-19.24^{+0.04}_{-0.06}$ & $49.00^{+0.11}_{-0.11}$ & $1.46^{+0.04}_{-0.04}$ & $0.29^{+0.02}_{-0.02}$ \\
		\hline
\end{tabular}
\end{center} 
\begin{tablenotes}
        \footnotesize
        \item[$*$] $^*$ In units of km$\cdot$s$^{-1}\cdot$Mpc$^{-1}$.
        \item[$^\star$] $^\star$With using $H_0 = 74.5^{+5.6}_{-6.1}$~km$\cdot$s$^{-1}\cdot$Mpc$^{-1}$ as a prior.
        \item[$^\dag$] $^\dag$ With 500 mock GRBs combined.
\end{tablenotes}
\end{table*}

\subsection{Distance indicators used to derive $d_l$ and $d_s$}\label{Sec:SC}

\begin{figure}
\includegraphics[scale=0.47]{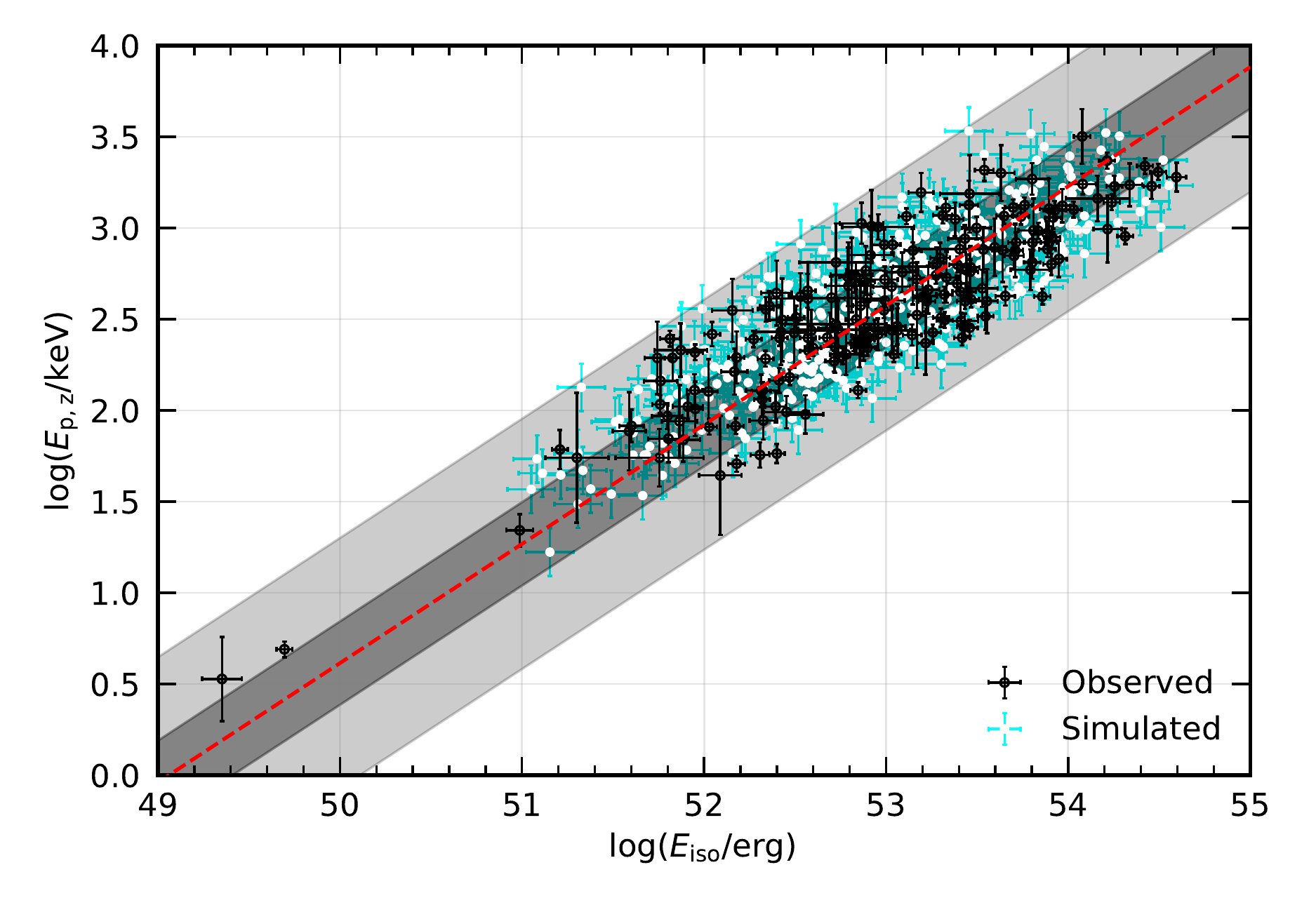}
\includegraphics[scale=0.235]{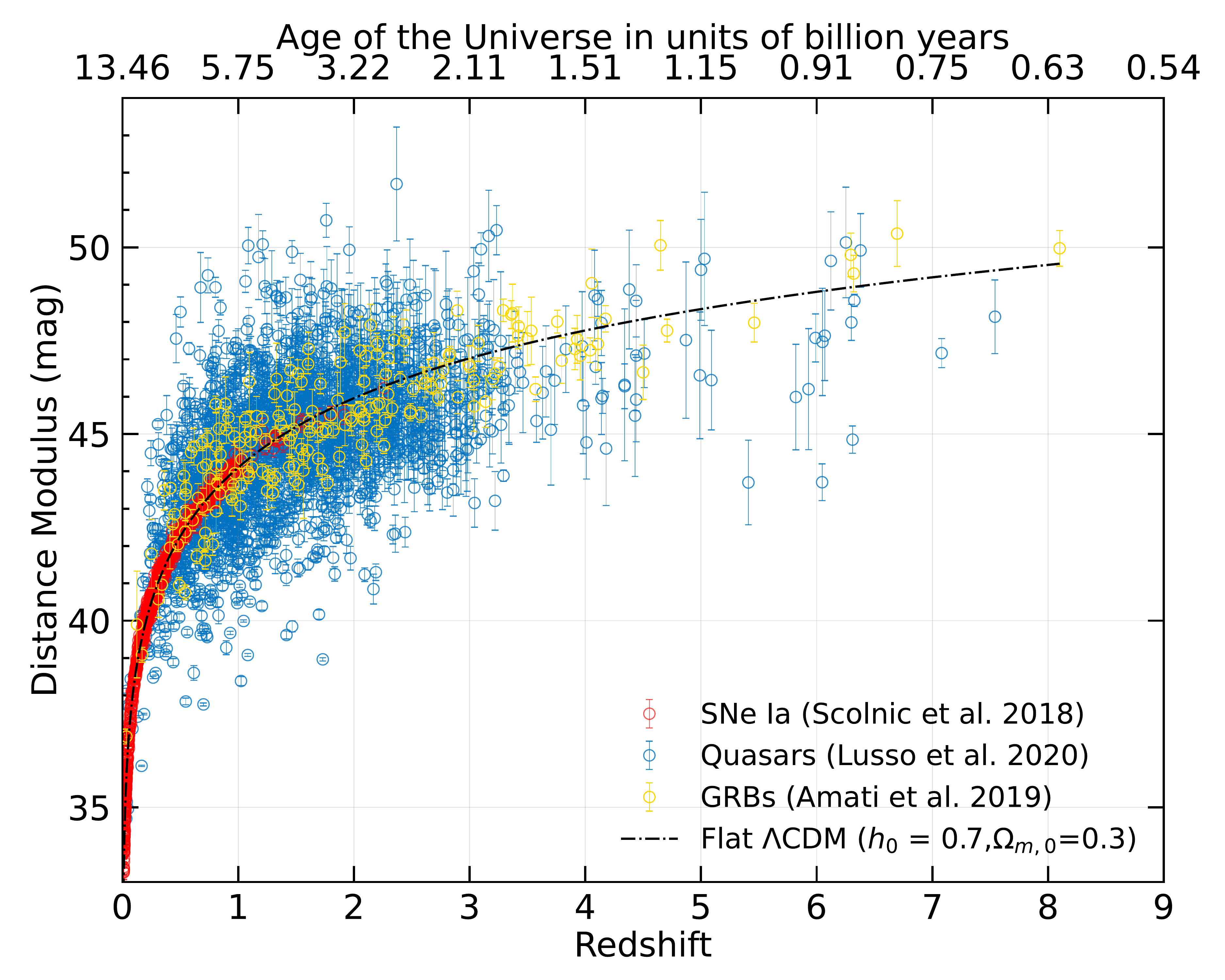}
\caption{Upper panel: The derived best-fitting $E_{\rm p,z} - E_{\rm iso}$ relation assuming a flat $\Lambda$CDM cosmology with taking $H_0= 70$~km$\cdot$s$^{-1}\cdot$Mpc$^{-1}$ and  $\Omega_{\rm m,0} = 0.3$ (red dashed line), 1$\delta_{\rm grb}$ (dark gray) and 3$\delta_{\rm grb}$ (light gray) confidence regions. The black symbols are the observed data and the cyan points denote the simulated (see section~\ref{Sec:simulate}). Lower panel: The Hubble diagrams of the standardizable candles implemented in this work, including SNe Ia (red), quasars(blue), and GRBs(orange). Note that we adopt the quasars with $z\le5$ for our analysis in this work (see Section~\ref{Sec:SC}). The term $h_0 = H_0/(100~\rm km\cdot s^{-1}\cdot Mpc^{-1})$.
}\label{MyFigA}
\end{figure}

{\bf Gamma-Ray Bursts.} 
We compile 193 redshift-known LGRBs from \cite{Amati2019MNRAS.486L..46A},  
and implement the Amati correlation as GRB distance indicators. 
For this GRB sample, $\sim70\%$ of them lie in the $0.5<z<2.5$ range with an average of $\bar z\sim1.9$.   
Similar to \cite{Amati2019MNRAS.486L..46A}, we write the Amati correlation as a 
power-law function of the cosmological rest-frame peak energy, 
$E_{\rm p,z}$, derived from the observer-frame one, $E_{\rm p,obs}$, via $E_{\rm p,z} = E_{\rm p,obs}(1+z)$, and the prompt $\gamma$-ray energy of LGRBs, 
$E_{\rm iso}=4\pi d^2_{\rm L}(z, \bm \Omega) S_{\rm bol}(1+z)^{-1}$, 
where $d_{\rm L}$ is the luminosity distance (in units of cm) in terms of an approximate parametrization
or a specific cosmology with free parameters $\bm \Omega$, 
$S_{\rm bol}$ is the observable bolometric integral flux (in units of erg$\cdot$cm$^{-1}$) within the energy band of observing instruments, 
and the factor $(1+z)$ is the correction of cosmological time dilation effect. 
Taking the logarithm, the $E_{\rm p,z} - E_{\rm iso}$ correlation can be modelled as
\begin{equation}\label{GRBindicator}
\log (E_{\rm iso}/{\rm erg}) = A +B\; \log (E_{\rm p,z}/\rm keV),
\end{equation}
with the constants $A$ and $B$ being the coefficient values of Amati correlation. 
Equation~(\ref{GRBindicator}) links the theoretical distance $d_{\rm L}(z, \bm \Omega)$ to the observable sets \{$z,E_{\rm p,z}, S_{\rm bol}$\} of LGRB population. 
Then the likelihood function is given such that
\begin{equation}\label{Eq:GRBlike}
\begin{split}
{\mathcal L}_{\rm grb} = \prod_{i=1}^{193}\frac{1}{\sqrt {2\pi(\sigma^2_{\log (S_{\rm bol}),i}+ B^2 \sigma^2_{\log (E_{\rm p,z}),i}+\delta^2_{\rm grb} )}}\\
\times \exp\left[-\frac{\Delta_{{\rm grb},i}^2}{2(\sigma^2_{\log (S_{\rm bol}),i}+ B^2 \sigma^2_{\log (E_{\rm p,z}), i}+\delta^2_{\rm grb})}\right], 
\end{split}
\end{equation}
and 
\begin{equation}
\Delta_{\rm grb}= \log (S_{\rm bol}) - A - B\;\log(E_{\rm p,z}) + \log(4\pi d^2_{\rm L}/(1+z)) ,
\end{equation}
Here, $\sigma_{\log E_{\rm p,z}}$ ($\sigma_{\log S_{\rm bol}}$) represents the error propagation (in logarithm) of the observed $E_{\rm p,z}$ ($S_{\rm bol}$), $\delta_{\rm grb}$ is a free parameter accounting for the intrinsic dispersion around the best-fitting relation of Equation~(\ref{GRBindicator}). 
Throughout we neglect the uncertainties of spectroscopically measured redshift and K-correction for taking into account the cosmological redshifting of the GRB spectra and the limited bandwidth of observing telescopes \citep{Bloom2001AJ....121.2879B}.

In the upper panel of Figure~\ref{MyFigA}, we present the best-fitting $E_{\rm p,z} - E_{\rm iso}$ relation (red dashed line) along with 1$\delta_{\rm grb}$ and 3$\delta_{\rm grb}$ confidence regions by assuming the flat $\Lambda$CDM as a fiducial cosmology ($H_0= 70$~km$\cdot$s$^{-1}\cdot$Mpc$^{-1}$, $\Omega_{\rm m,0} = 0.3$, the same as below), where $A =49.06\pm 0.32$, $B = 1.53\pm 0.12$, and $\delta_{\rm grb} = 0.35\pm 0.04 $~dex. 
The derived GRB Hubble diagram is plotted in the lower panel (see orange points). 
One can see that the GRBs distance moduli are consistent with the SN Ia ones at low redshifts \footnote{The absolute magnitude of all SN Ia objects is fixed to -19.35~mag, which corresponds to $H_0 = 70$~km$\cdot$s$^{-1}\cdot$Mpc$^{-1}$, see \cite{Scolnic2018ApJ}. }, albeit with an intrinsic dispersion of around 0.35~\emph{dex} in logarithmic space.  
Besides, \cite{Demianski2021MNRAS.tmp.1707D} parametrized the redshift evolution of Amati parameters in the form of a power-law function and showed that there is no significant trend systematic with redshift for the observables, coefficients, as well as  $\delta_{\rm grb}$, by cross-comparing the calibrations in different redshift bins and observing instruments using a local regression technique. Therefore, we can safely use this sample of LGRBs as distance indicators.

{\bf Pantheon Supernovae Ia.} 
The \emph{Pantheon} compilation released the current most complete SN Ia sample, 
which consists of 1048 objects covering the redshift range of $0.01\le z \le 2.3$ \citep{Scolnic2018ApJ}. As shown in the lower panel of Figure~\ref{MyFigA}, most of the SN Ia events ($\sim 80\%$) are located in the redshift range  $0.2<z<0.6$, and only 6 of them are at $z>1.4$, much insufficient to be used to cover the redshift range of current strong lensing systems. 
The observed SN Ia distance moduli (red points in this panel) are calculated from the Spectral Adaptive Light curve Template 2 (SALT2) light-curve fit parameters 
with using a modified version of the Tripp formula \citep{Tripp1998A&A...331..815T}:
\begin{equation}
\label{eq:SNmu}
\hat{\mu}_{\rm obs}=m_{\rm B} -M_{\rm B}+\alpha \cdot X -\eta \cdot C+\Delta_{\rm M}+\Delta_{\rm B},
\end{equation}
where $m_{\rm B}$ is the B-band apparent magnitude, 
$X$ is the stretch factor, $C$ is the color parameter, 
$\alpha$ and $\eta$ are the nuisance coefficients of the stretch- and color-luminosity relations, respectively, 
and $M_{\rm B}$ is another nuisance parameter that represents the B-band absolute peak magnitude of a fiducial SNe Ia with $X$ = 0 and $C$ = 0. 
Furthermore, $\Delta_{\rm M}$ and $\Delta_{\rm B}$ are the distance corrections based on the host-galaxy mass and from various biases predicted from simulations, respectively. 
For the Pantheon SNe Ia, the values of $\alpha$ and $\eta$ are calibrated by using 
the Bayesian Estimation Applied to Multiple Species with Bias Corrections (BBC) method \citep{Kessler2017ApJ...836...56K}, 
whistle correcting those expected biases (See \citealt{Marriner2011ApJ...740...72M} for more details.). 
With the BBC technique, \cite{Scolnic2018ApJ} reported the apparent magnitudes with bias corrections ($m_{\rm corr}$) of all SNe Ia. 
Then the observed distance modulus is given by directly subtracting $M_{\rm B}$ from $m_{\rm corr}$, i.e., $\hat{\mu}_{\rm obs}=m_{\rm corr}-M_{\rm B}$. 
Using the vector of SN distance residuals $\Delta${\bf m}$=\hat{\bm \mu}_{\rm obs}-{\hat {\bm \mu}}_{\rm model}$, the likelihood function of the model fit is given by 
\begin{equation}
-2 \ln ({\mathcal L}_{\rm sn}) = \Delta {\bf m}^{T} \cdot  {\bf C}^{-1} \cdot \Delta {\bf m}.
\end{equation}
Here, ${\bf C}$ is the uncertainty matrix (including both statistical and systematic uncertainties), $\hat{\bm \mu}_{\rm obs}$ and ${\hat {\bm \mu}}_{\rm model}$ are respectively the observed and model vector of SN distance moduli with $\hat \mu_{\rm model}=5 \log [d_{\rm L}({z, \bm \Omega})/(10~{\rm pc})]$. 

Actually, in our method SN Ia data can be anchored near their high-redshift end in combination with  $D_{\Delta t}$ measurements from TDGLs \citep{Taubenberger2019A&A...628L...7T, Wong2020MNRAS.tmp.1661W}, thus we do not take a fiducial value of $H_0$ in SN Ia distances in our fitting.

{\bf Quasar Sample.} 
Quasars are of high luminosity emerging at galaxy scales and have served as a new cosmological probe of the high-redshift universe  \citep{Risaliti2019NatAs...3..272R, Lusso2020A&A...642A.150L}.   
Recently, \cite{Lusso2020A&A...642A.150L} reported a new catalogue of 2421 optically selected quasars with the measurements of spectroscopic redshifts and X-ray observations from either Chandra or XMM-Newton. This high-quality sample enables set up the Hubble diagram up to redshift $z\sim 7.5$ by using the non-linear phenomenological relation between the X-ray and ultraviolet luminosities of quasars \citep{Avni1982ApJ...262L..17A, Lusso2017A&A...602A..79L}, written logarithmically as
\begin{equation}\label{Eq:LXLUV}
\log L_{\rm X}= \beta+\gamma \log L_{\rm UV},
\end{equation}
where $L_{\rm X}$ and $L_{\rm UV}$ correspond to the monochromatic luminosities at 2~keV and 2500~${\rm \AA}$, respectively. 
Then the distance indicator of quasars can be re-expressed as a function of the UV and X-ray fluxes (both in units of erg\,cm$^{-2}$\,s$^{-1}$\,Hz$^{-1}$), $F_{\rm UV}$ and $F_{\rm X}$, luminosity distance, $d_{\rm L}$ at redshift $z$, i.e.,
\begin{equation}\label{Eq:FXFUV}
\log F_{\rm X} = \beta+\gamma \log F_{\rm UV}+(\gamma-1) \log 4\pi d^2_{\rm L}(z).
\end{equation}
We can obtain the quasar distances by directly fitting the relation between $F_{\rm X}$ and $F_{\rm UV}$ as a function of the slope $\gamma$ and intercept $\beta$. The redshift dependence of $\gamma$ and $\beta$ has been evaluated in a cosmology-independent way by using sub-samples in narrow redshift bins. 
With this approach,  \cite{Lusso2020A&A...642A.150L} verified that both $\gamma$ and $\beta$ do not show significant trend with redshift at $z\le5$. 
We thus use 2404 quasars with redshift $z\leq5$ in the following analyses. 
It should be noted that the intercept $\beta$ is very difficult to be determined due to the lack of confident physical interpretations for the $L_{\rm X}-L_{\rm UV}$ relation as well as its intrinsic dispersion, $\delta_{\rm qso}$. Therefore, we set $\gamma$, $\beta$, and $\delta_{\rm qso}$ as free parameters to be marginalized over in our fitting, which can also partly alleviate the problem of the mismatch in the shape of Hubble diagrams of quasars and other probes (e.g., the SNe Ia). 
The quasar likelihood function is given by
\begin{equation}
\begin{split}
{\mathcal L}_{\rm qso}= \prod_{i=1}^{2404}\frac{1}{\sqrt{2\pi(\sigma^2_{\log (F_{\rm X}),i}+ \gamma^2 \sigma^2_{\log (F_{\rm UV}),i}+\delta^2_{\rm qso} )}}\\
\times\exp\left[-\frac{\Delta_{{\rm qso},i}^2}{2(\sigma^2_{\log (F_{\rm X}),i}+ \gamma^2 \sigma^2_{\log (F_{\rm UV}),i}+\delta^2_{\rm qso})}\right], 
\end{split}
\end{equation}
and 
\begin{equation}
\Delta_{\rm qso}=\log (F_{\rm X,30}) - \beta' - \gamma\; \log (F_{\rm UV, 27.5}) - 2(\gamma-1)\log (d_{\rm L,28.5}(z, {\bm \Omega})), 
\end{equation}
where $\beta'= \beta +(\gamma-1)\log(4\pi)$, $F_{\rm X}$ and $F_{\rm UV}$ are normalised (in logarithm) to 30 and 27.5, respectively.

The lower panel of Figure~\ref{MyFigA} also displays the quasar Hubble diagram (blue symbols), which is derived by fitting the data with fifth-order logarithmic polynomials \citep{Lusso2020A&A...642A.150L}. 
We can see that quasars exhibit a relatively large dispersion and appear to deviate from the flat $\Lambda$CDM beyond $z\sim2$ (see \citealt{Lusso2020A&A...642A.150L} for more details). 
We thus focus on the cosmographic fits with the combined $D_{\Delta t}$ of TDGLs, SN Ia, and GRB data. Nevertheless, the combined results with quasars can be employed for comparison and validation in our model-independent measurements.

\section{Results and Discussions}\label{Sec:sec4}
\subsection{Current observational constraints}\label{Sec:results}

\begin{figure*}
\includegraphics[width=4.5in]{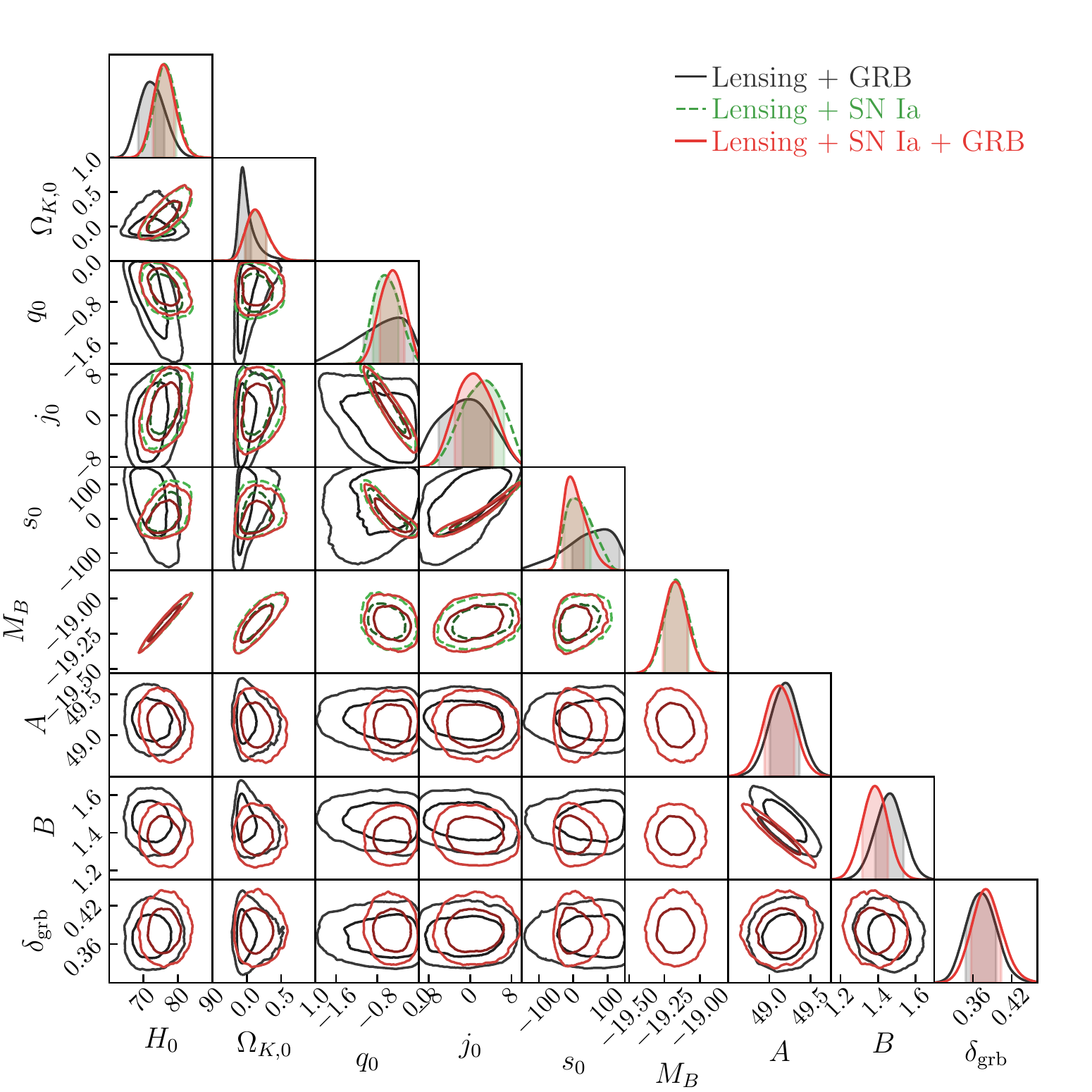}
\includegraphics[width=4.5in]{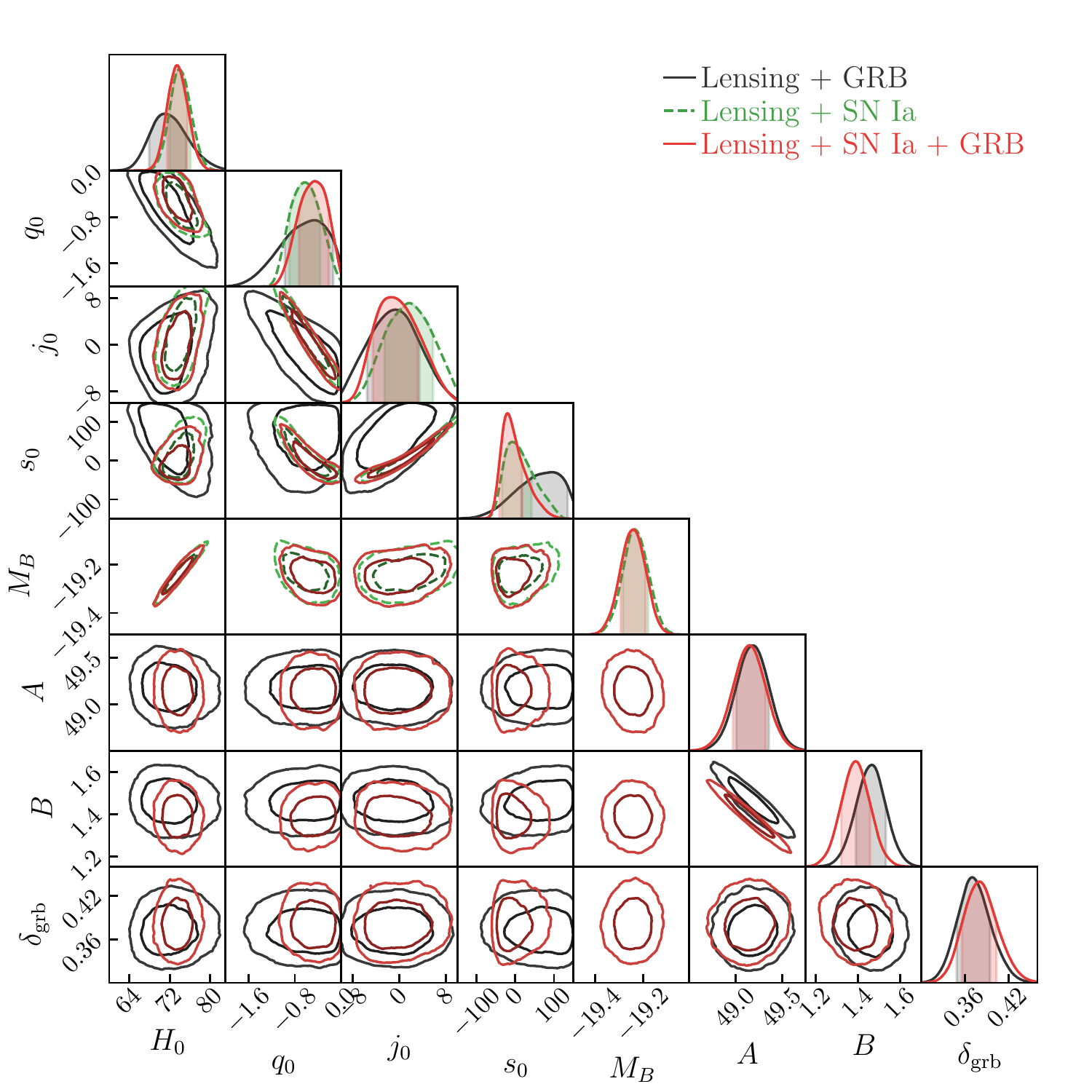}
\caption{Marginalized one- and two-dimensional posterior probability distributions (and contours) at 1$\sigma$ and 2$\sigma$ credibilities produced by dynamical nested sampling with the combination of TDGLs, SNe Ia, and GRBs in general cosmography with a curvature component (upper panel) and with taking a flat universe prior (lower panel).  
}\label{MyFigB}
\end{figure*}

\begin{figure*}
\includegraphics[width=4.5in]{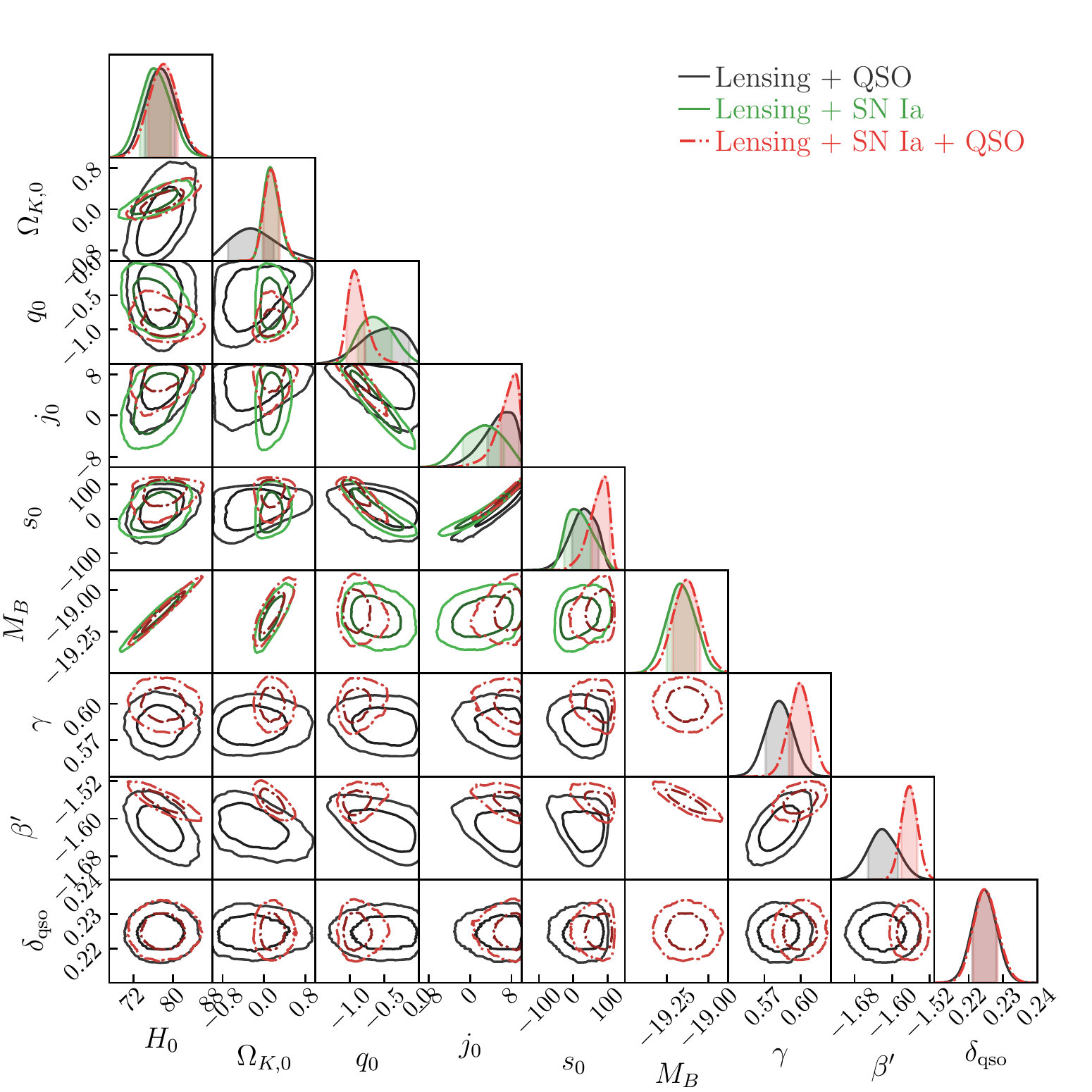}
\includegraphics[width=4.5in]{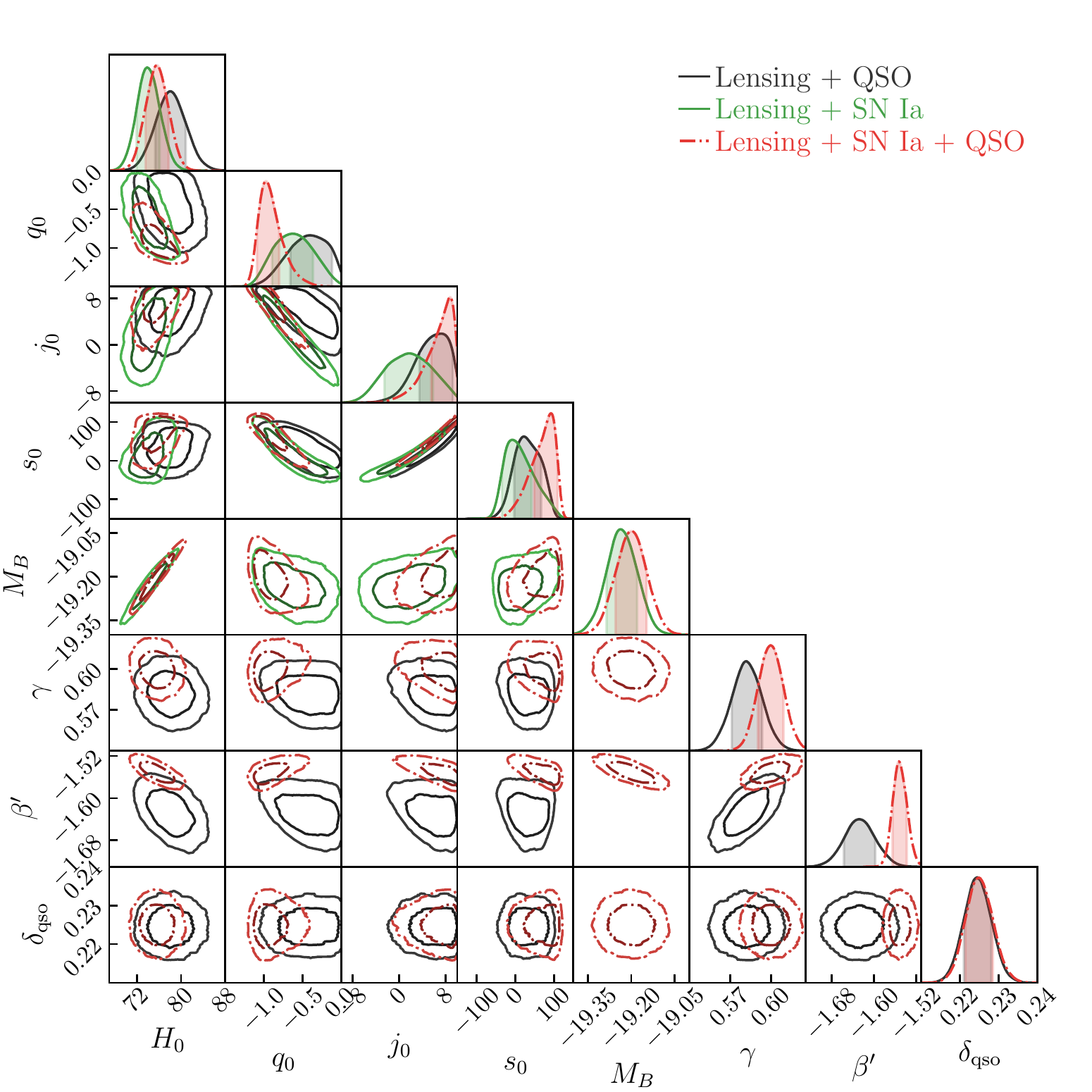}
\caption{Same as in Figure~\ref{MyFigB}, but with the combination of TDGLs, SNe Ia, and quasars. 
}\label{MyFigC}
\end{figure*}

We first focus on the cosmographic fits by combining TDGLs, SNe Ia, and GRBs. For all the free parameters, we use uninformative priors, in particular, for the cosmographic parameters, we adopt the priors used by \cite{Li2020MNRAS.491.4960L}. 
In Table~\ref{MyTabA}, we give the priors and detailed descriptions of each free parameter to be constrained.  
Table~\ref{MyTabB} tabulates the best-fitting results and 1$\sigma$ credible intervals produced from a random sampling of the Bayesian formulae using the dynamic nested sampling code. We display the 1$\sigma$ and 2$\sigma$ marginalized posterior probability distributions (contours) in the \emph{cornerplots} of Figure~\ref{MyFigB}. 
Combining the lensing and low-redshift SNe Ia yields $H_0= 75.8^{+3.7}_{-2.5}$~km s$^{-1}$Mpc$^{-1}$ and $\Omega_{K,0}= -0.12^{+0.16}_{-0.14}$, with smaller statistical uncertainties compared to previous results \citep{Collett2019PhRvL.123w1101C}, which can be attributed to the fact that the constraints of $H_0$ and $\Omega_{K,0}$ are contributed by both the $D_{\Delta t}$ and SC distances. This also suggests that the cosmographic parametrizations with a spatial curvature component can, in some degrees, consolidate the direct measurements of $H_0$ and $\Omega_{K,0}$ via the DSR. 
Taking a flat-universe prior, we obtain $H_0= 73.8^{+2.3}_{-1.8}$~km s$^{-1}$Mpc$^{-1}$, significantly improved by 38\%. 
From the upper panel of Figure~\ref{MyFigB}, we notice that the degeneracy between $M_B$ and $H_0$ in SN Ia distances still exists due to the limited constraints on $H_0$ by current $D_{\Delta t}$ data, which degrades the determinations of $H_0$ and then $\Omega_{K,0}$ when combining SNe Ia. 
To reduce such degeneracy effects, we utilize the prior of $H_0$ with the most recently released measurement by SH0ES team ($H_0 = 73.30\pm 1.04$~km s$^{-1}$Mpc$^{-1}$; \citealt{Riess2021ApJ...908L...6R}), and find that the $\Delta \Omega_{K,0}$ falls by $\sim 40\%$. 
From the Lensing+GRB combination, we obtain $H_0= 71.5^{+4.4}_{-3.0}$~km s$^{-1}$Mpc$^{-1}$ and $\Omega_{K,0}= -0.07^{+0.13}_{-0.06}$, providing a 6\% precision measurement of $H_0$ and favoring a flat universe with $\Delta \Omega_{K,0} \approx 0.1$. 

The uncertainties of combining Lensing+GRB mainly come from the poor constraints of cosmographic parameters due to the large dispersion of current GRB data.  
In the case of a flat universe, the error bar of $H_0$ is slightly narrowed (by $\sim 5\%$), and the cosmographic parameters are also moderately improved. 
In addition, the derived values of deceleration, jerk, and snap parameters are in agreement with a present accelerating universe in $\Lambda$CDM cosmology, which approximately corresponds to $q_0 \approx -0.55$ ($\Omega_{\rm m,0} = 0.3$), $j_0 = 1$, and $s_0 \approx 0$ \citep{Capozziello2013Galax...1..216C}, although $s_0$ is poorly constrained.

SN Ia data possess much better qualities at relatively low redshifts relative to GRBs (and quasars), and thus can be implemented as references to calibrate the GRB (quasar) distances in the same redshift range. 
Adding the SN Ia data into the Lensing+GRBs, we find that the uncertainty of $H_0$ is narrowed by 25\% relative to the value determined by Lensing+GRBs and the $H_0-M_B$ degeneracy can not be improved. 
The constraint of $\Omega_{K,0}$ is not improved, which could be the result of the effect of $H_0-M_B$ degeneracy in SN Ia distances and the mismatch between SN Ia and GRB Hubble diagram at $z\gtrsim 1.4$\footnote{This can be seen from the constrained values of the slope $B$ and snap parameter between Lensing+GRB and Lensing+SN Ia+GRB, see Table~\ref{MyTabB}.}. 
Such degeneracy effects can be seen from the result by taking the SH0ES prior on $H_0$, which reduces $\Delta \Omega_{K,0}$ by 39\%. 
In our prospects below, we will show the effects of the uncertainties regarding to GRB distances on the measurements of $H_0$ and $\Omega_{K,0}$ combining a larger sample of realistic GRBs produced based on a Monte-Carlo simulation.

Figure~\ref{MyFigC} presents the results by combing Lensing, SNe Ia, and quasars; see also Table~\ref{MyTabB}. 
In the case of Lensing+QSO, we obtain a relatively higher value of $H_0$ with $\Delta H_0\approx 3$~km s$^{-1}$Mpc$^{-1}$; the spatial curvature density is loosely constrained with $\Delta \Omega_{K,0}\sim 0.5$. Compared to Lensing+GRB, the looser constraint on $\Omega_{K,0}$ comes from that the quasar distances ($z\lesssim 5$) have a larger dispersion and  larger error bars at high redshifts (e.g., $z\gtrsim 2.5$; see the lower panel in Figure~\ref{MyFigA}), which results in larger uncertainties in the measurements of $d_l$ and $d_s$ and then the $\Omega_{K,0}$. Also, GRB distances have a wider available redshift coverage than the current quasars, which can provide more meaningful constraints. 
Taking a flat-universe prior gives only a slightly improved value of $H_0$. 
When adding SN Ia data into Lensing+QSO, we obtain only a slight improvement on the limit of $H_0$, whereas the $\Delta \Omega_{K,0}$ can be reduced by a factor of 3. 
While the cosmographic parameters are more constraining relative to Lensing+SN Ia/QSO, their values deviate from that of $\Lambda$CDM cosmology at beyond 1$\sigma$ credible intervals. 
Meanwhile, compared to the results combining GRB distances above, the outcomes with QSOs also suffer from the $M_B-H_0$ degeneracy in fitting SN Ia distances and the mismatch between SN Ia and QSO Hubble diagrams. 
Therefore, one should be careful when combining quasar data as complementary distance indicators. 
The flat-universe assumption provides a 21\% reduction for the error bar of $H_0$. 
When imposing the SH0ES prior on $H_0$, we obtained a precision similar to the SN Ia and GRB yields.

When adding GRBs to Lensing+SN Ia+QSO, we find no systematic deviation between QSO and GRB distances. There is no significant improvement in our results ($H_0 = 77.4^{+2.9}_{-3.0}$~km s$^{-1}$Mpc$^{-1}$ and $\Omega_{K,0} = 0.13^{+0.17}_{-0.13}$) compared to Lensing+SNIa+QSO and/or Lensing+SNIa+GRB. The result could be due to the mismatch between the SN Ia and GRB/QSO distances, which is a sign of some systematic errors in QSO/GRB distance calibration as using the current $D_{\Delta t}$ measurements for the anchoring along with using SN Ia distances as a reference. 
Such uncertainties can be further analyzed by using more high-quality time-delay lensing systems, as which ``govern'' the precision and accuracy in restricting the $H_0$. Also, high-quality and large-statistics SNIa/GRB data sets are needed to find out the underlying systematical effects.

\begin{figure}
\begin{center}
\includegraphics[width=2.2in]{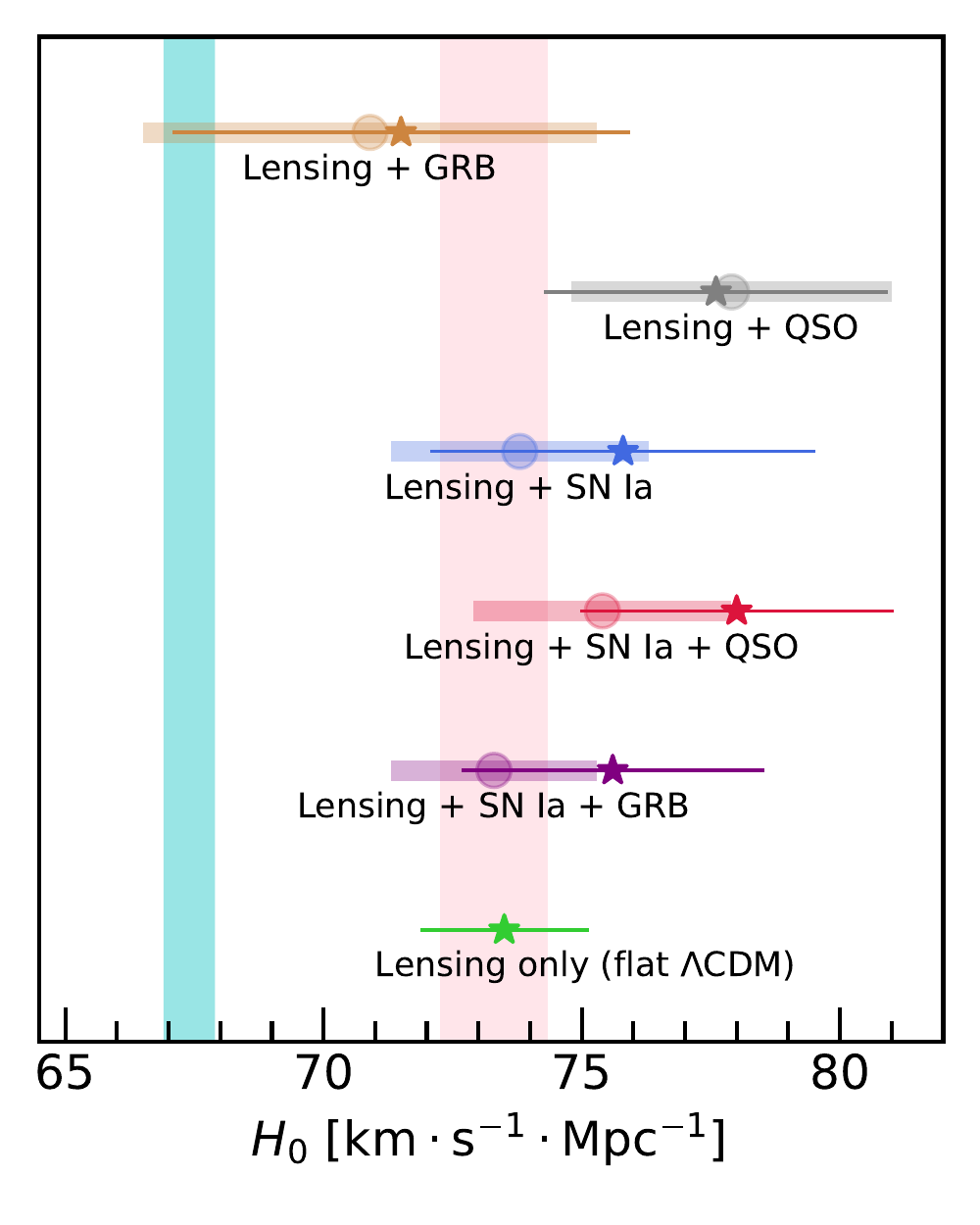}
\end{center}
\caption{The present values of Hubble parameter determined with different data combinations. 
The thin and shaded lines denotes the 1$\sigma$ results with taking a uninformative prior and  flat-universe prior on $\Omega_{K,0}$, respectively.  
The cyan shaded region denotes the CMB (early-universe) measurement in flat $\Lambda$CDM cosmology ($H_0 = 67.4\pm 0.5$~km$\cdot$s$^{-1}\cdot$Mpc$^{-1}$, \citealt{Planck2018resultsVI}) and the pink shaded region shows the (late-universe) determination by local distance ladder ($H_0 = 73.30\pm 1.04$~km$\cdot$s$^{-1}\cdot$Mpc$^{-1}$, \citealt{Riess2021ApJ...908L...6R}). 
}\label{MyFigD}
\end{figure}

In Figure~\ref{MyFigD}, we also display the measurements of $H_0$ with different data combinations. The value with combining Lensing+GRB is between the CMB and SH0ES results with a precision of $\sim$6\% (see Table~\ref{MyTabB}). 
The Lensing+SN Ia combination results in a value favoured by the SH0ES measurement (pink shaded region) with a $\sim$3\% precision in a flat universe. 
However, the combinations of quasars give a relatively high expansion rate. 
When combining with SNe Ia, the values coincide with the SH0ES measurement as well as the lensing-only result in a flat $\Lambda$CDM cosmology ($73.5\pm 1.6$~km$\cdot$s$^{-1}\cdot$Mpc$^{-1}$ from this work, see the green symbol) at 1$\sigma$ credible regions. 

In our method, the constraints of $H_0$ and $\Omega_{K,0}$ rely on the $d_l$ and $d_s$ provided by anchoring the pairs of SC objects on the $D_{\Delta t}$ along with their wide redshift coverage. 
The measurement of $D_{\Delta t}$ depends on the assumption in lens modelling, which dominates the source of systematic uncertainty in measuring $H_0$ in time-delay cosmography. \cite{Birrer2020A&A...643A.165B} considered the mass-sheet transformations and obtained $H_0 = 74.5^{+5.6}_{-6.1}$~km s$^{-1}$Mpc$^{-1}$. 
We also use this value as a prior to see the impact of the systematic effects arising from lens modelling on the limit of $\Omega_{K,0}$. The results are presented in Table~\ref{MyTabB}. 
One can see that such effects do not significantly influence the bound of $\Omega_{K,0}$.

\subsection{\bf Future prospects} \label{Sec:simulate}
As shown above, the determinations of $H_0$ and $\Omega_{K,0}$ via the DSR suffer from the lack of low-redshift GRB observations and the large intrinsic dispersion in the GRB Hubble diagram. 
Such uncertainties, as well as the effect of $M_B-H_0$ degeneracy in fitting SN Ia distances, can be improved with larger statistics of GRB data, although the measurement of $H_0$ is mainly determined by the lensing $D_{\Delta t}$ measurements. 
Here we perform a Monte-Carlo simulation to investigate the constraints on the cosmographic and spatial curvature parameters by combining the current lensing, Pantheon SN Ia sample, and future GRB observations of the Amati quantities.
Our simulation procedure for producing the $z, E_{\rm p,z}$, and $S_{\rm bol}$ of each GRB event is as follows. 
First, we fit the gamma distribution function to the current spectroscopic redshift data set, and randomly sample the redshifts in the range $0.001< z< 2.5$, which well covers the current lensing systems. 
Second, we fit the gaussian function to the $\log(E_{\rm p,z}/\rm keV)$ data, which gives a mean 2.57 and width 0.45, and $\log(E_{\rm p,z}/\rm keV)$ of each mock GRB is randomly drawn with subject to such a density function. 
Third, we produce the mock $E_{\rm iso}$ samples using the best-fitting Amati correlation in Equation~(\ref{GRBindicator}) with considering an intrinsic dispersion $\delta_{\rm grb} = 0.3$~dex, which are determined in a flat $\Lambda$CDM cosmology (see Section~\ref{Sec:SC}). 
The jet collimated corrections of the GRB prompt $\gamma$-ray energy are neglected, despite the arguments that the jet opening angles could be narrow as below 10~degrees (e.g., \citealt{Goldstein2016ApJ...818...18G}). 
Finally, we calculate the mock $S_{\rm bol}$ data from the simulated $E_{\rm iso}$ by evaluating the luminosity distances assuming a flat $\Lambda$CDM model and select the GRB events with a moderate detector threshold of $1\times 10^{-7}$~erg$\;$cm$^{-2}$ (e.g., at the energy band 20-2000~keV). 
We specify a common statistical uncertainty of 30\% for the simulated samples of $\{E_{\rm p,z}, S_{\rm bol}\}$. 
By running the Monte-Carlo simulation, we produce 500 realistic GRB events for our following analysis\footnote{The expected number of GRBs observed at the design sensitivities of current (Swift, \citealt{Gehrels2004ApJ...611.1005G}; and Fermi, \citealt{Meegan2009ApJ...702..791M}) and future (SVOM, \citealt{Gotz2009AIPC.1133...25G}; LOFT, \citealt{Feroci2012ExA....34..415F}; and THUSEUS, \citealt{Amati2018AdSpR..62..191A}) missions dedicated to observing GRBs in the coming decade’s operations can reach $\sim600$. }, which are shown in the upper panel of Figure~\ref{MyFigA} (cyan pints). 

\begin{figure*}
\includegraphics[width=4.5in]{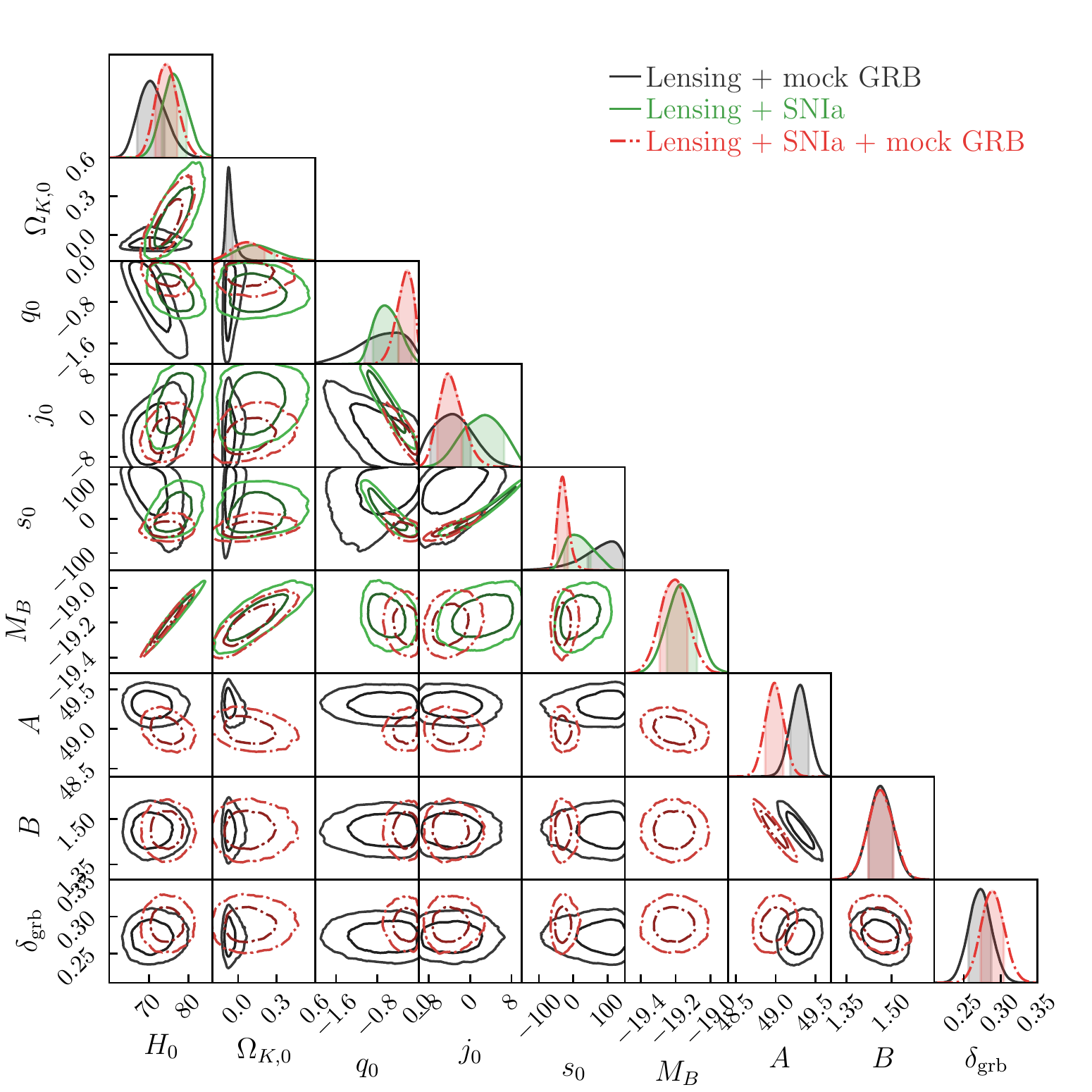}
\includegraphics[width=4.5in]{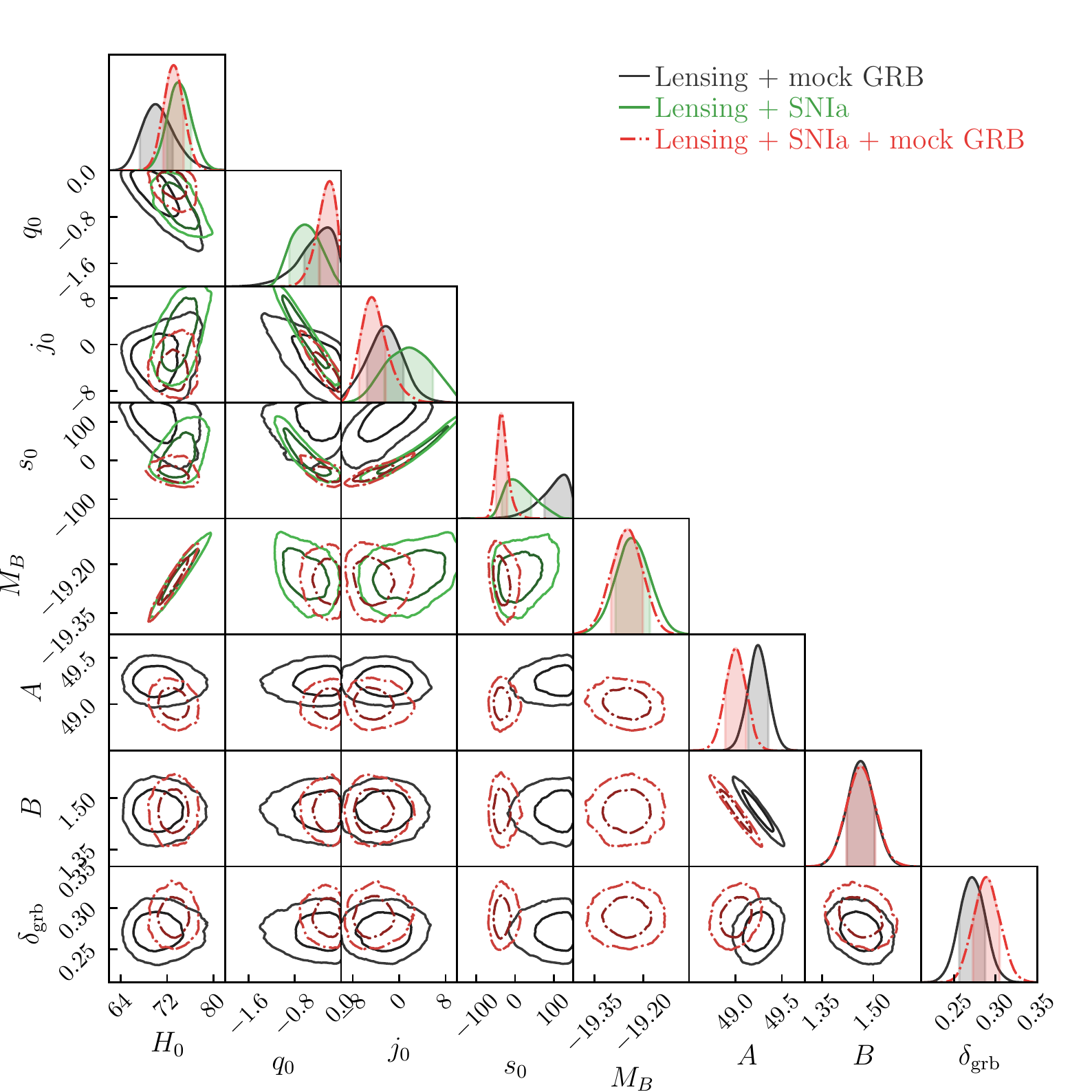}
\caption{Same as in Figure~\ref{MyFigB}, but with the combination of TDGLs, SNe Ia, and 500 mock GRBs.
}\label{MyFigE}
\end{figure*}

Figure~\ref{MyFigE} presents the results by combining the 500 mock GRB events. 
The 1$\sigma$ credible intervals are also summarized in Table~\ref{MyTabB}. 
We find that the restrictions of cosmographic parameters are more stringent compared to that with the real data and the $\Delta \Omega_{K,0}$ can be reduced to $\sim 0.02$ by using Lensing+mock GRB. 
The case of using a flat-universe prior further reduces the error bar of $H_0$ by 14\% and results in a 4\% precision. 
When adding the mock GRBs into Lensing+SN Ia, the $M_B-H_0$ degeneracy can be moderately eliminated (the contours in the $M_B-H_0$ plane are narrowed and the correlation between $M_B$ and $H_0$ tends to be weaker) and both the uncertainties of $H_0$ and $\Omega_{K,0}$ are narrowed compared to Lensing+SN Ia and/or Lensing+SN Ia+GRB. 
Using the prior of $H_0$ informed by SH0ES, we can further eliminate the $M_B-H_0$ degeneracy in SN Ia distances and the result of $\Omega_{K,0}$ is more constraining.  
In the assumption of a flat universe, the precision of $H_0$ can reach $\sim 2$\%.

\section{Summary}\label{Sec:sec5}

We have provided new independent constraints on the $H_0$ and $\Omega_{K,0}$ combining the $D_{\Delta t}$ measurements from TDGLs and GRB luminosity distances, 
only based on the validity of the FLRW metric and geometrical optics. 
This method benefits from the wide redshift coverage of the absolute distances of lensing systems by using the high-redshift LGRBs, taking the full advantage of  $D_{\Delta t}$ measurements in determining $H_0$. 
We use the Amati correlation with 193 redshift-known LGRBs to establish the GRB Hubble diagram up to $z\sim 8.1$, much higher than that of SNe Ia or quasars. 
The $H_0$ and $\Omega_{K,0}$ are determined by fitting the $D_{\Delta t}$ and GRB distance indicators through the DSR, in which the GRB distances are parametrized using the general cosmographic parameters together with a curvature component. 
Thus the limitations of $H_0$ and $\Omega_{K,0}$ are contributed by both the combined $D_{\Delta t}$ and SC distances. 

We present our main findings below. 
With the combination of Lensing+GRB, we obtain $H_0= 71.5^{+4.4}_{-3.0}$~km s$^{-1}$Mpc$^{-1}$ and $\Omega_{K,0}= -0.07^{+0.13}_{-0.06}$. 
In a flat universe, the $H_0$ is measured to be $H_0= 70.9^{+4.2}_{-2.9}$~km s$^{-1}$Mpc$^{-1}$, with a 6\% precision. 
The SN Ia distances can be anchored near their high redshift end in the combination of Lensing+SN Ia, in which the constraints of $H_0$ and $\Omega_{K,0}$ suffer from the $M_B-H_0$ degeneracy in SN Ia distances due to the limited precision of $H_0$ provided by $D_{\Delta t}$ measurements of only six TDGLs. 
In the combination of Lensing+SN Ia+GRB, current GRB data can neither eliminate the $M_B-H_0$ degeneracy when fitting to SN Ia distances nor effectively improve the constraint on $H_0$; the mismatch between the SN Ia and GRB Hubble diagrams due to some systematic errors in the GRB distance calibration contributes to a part of uncertainty in restricting $\Omega_{K,0}$. 
We performed a Monte-carlo simulation to produce a larger sample of realistic GRBs, and showed that current Lensing+mock GRB could yield inspiring improvements in measuring $H_0$ (with 4\% precision) and $\Omega_{K,0}$ (with an uncertainty of $\sim 0.02$). 
Future increment of GRB observations could help eliminate the $M_B-H_0$ degeneracy in SN Ia data and reconcile the SN Ia and GRB Hubble diagrams in the combination of Lensing+SN Ia+GRB, providing more constraining results of $H_0$ and $\Omega_{K,0}$. 
In addition, we only include one double-source-plane strong lensing system in our sample, which provides an accurate measurement of the cosmological scaling factor that is sensitive to the value of $\Omega_{K,0}$. 
The inclusion of more such events will further improve the restrictions on $\Omega_{K,0}$. 
In contrast, the resultant values of $\Omega_{K,0}$ by using quasars are not very constraining compared to SNe Ia and/or GRBs in conjunction with TDGLs, although they have relatively large statistics. 
In the combinations with current lensing, SNe Ia, and GRB data sets, we find no evidence of significant deviation from a zero spatial curvature or an accelerating universe, and the values of $H_0$ coincide with the result of local distance ladder with the best accuracy $\Delta H_0\approx 2.0$~km s$^{-1}$Mpc$^{-1}$ (3\% precision) in a flat universe.

To date, various solutions for the $H_0$ tension between the CMB and local measurements (e.g., any new pre-combination physics or the modification of the post-combination expansion history of the Universe) are unlikely to be successful (see e.g., \citealt{Knox2020PhRvD.101d3533K} and reference therein). 
In addition, various independent measurements might suffer from some unknown systematic effects (see a recent review by \citealt{Di2021CQGra..38o3001D}). 
More precise, model-independent measurements of $H_0$ and $\Omega_{K,0}$ via the DSR with the increases in the number of time-delay lensing systems (in which more sophisticated lens modelling are expected) and in the statistics of reliable standardizable candles are admired for verifying and providing some clues to the solution of the $H_0$ tension. 
Such measurements also give clues to the nature of dark energy and provide a consistency test of the classical FLRW metric.

\section*{Acknowledgements}
We thank the referee for very helpful comments and suggestions. We are grateful to L. Amati for providing the GRB data and Elisabeta Lusso for sharing their latest released quasar sample. 
Z.-H. Zhu is supported by the National Natural Science Foundation of China under Grants Nos. 12021003, 11920101003 and 11633001, and the Strategic Priority Research Program of the Chinese Academy of Sciences, Grant No. XDB23000000. 
J.-J. Wei is supported by the NSFC under grant No. U1831122, the Key Research Program of Frontier Sciences (grant No. ZDBS-LY-7014) of Chinese Academy of Sciences, and the Natural Science Foundation of Jiangsu Province (grant No. BK20221562). E.-W. Liang is supported by the National Natural Science Foundation of China (grant Nos. 12133003).
Z.-C. Chen is supported by the National Natural Science Foundation of China (Grant No.~12247176) and the China Postdoctoral Science Foundation Fellowship No.~2022M710429.

\section*{Data Availability}
The data of strong lensing systems we use are available at \url{http://www.h0licow.org}. The GRB and quasar sample are shared on reasonable requests to \cite{Amati2019MNRAS.486L..46A} and \cite{Lusso2020A&A...642A.150L}, respectively. 
The PYTHON/JUPYTER notebook to reproduce the results of this article will be shared on reasonable request to the corresponding author.


\bibliographystyle{mnras}
\bibliography{main}

\appendix
\section{Appendix}
In this appendix, we present the formulas for the parametrizations of the cosmic distances used in the general cosmography.

The leading-order cosmographic parameters are defined as the scale factor deviations in terms of cosmic time:

\begin{equation}\label{Eq:CosmParm}
H(t) \equiv \frac{1}{a}\frac{da}{dt}, q(t) \equiv -\frac{1}{a H^2}\frac{d^2a}{dt^2}, j(t) \equiv \frac{1}{a H^3}\frac{d^3a}{dt^3}, s(t) \equiv \frac{1}{a H^4}\frac{d^4a}{dt^4}, ... ,
\end{equation}
which are respectively referred to the Hubble, deceleration, jerk, and snap parameters, etc. 
With these definitions, we investigate the cosmic expansion with expanding the scale factor into a Taylor series around the present time $t_0$ ($z=0$), 
\begin{equation}\label{Eq:scalefactor}
\begin{split}
\frac{a(t)}{a(t_0)}= 1+H_0(t-t_0) -\frac{1}{2}q_0 H_0^2(t-t_0)^2+\frac{1}{6}j_0 H_0^3(t-t_0)^3 
       \\ +\frac{1}{24} s_0 H_0^4(t-t_0)^4 +... ]. 
\end{split}
\end{equation}
Using the FLRW geometry and the definition of redshift $z=1/a(t)-1$ with simply assuming $a(t_0)=1$, we then obtain the dimensionless co-moving distance as a function of redshift with the fourth-order cosmographic parameters, 
\begin{equation}\label{Eq:dcz}
\begin{aligned}
d(z)=z-\frac{1}{2}(1+q_0)z^2 +\frac{1}{6}(2+4q_0+3q_0^2-j0+ \Omega_{K,0})z^3 \\
-\frac{1}{24}[6+18q_0-27q_0^2-15q_0^3-(9+10q_0)j_0\\
- s_0+(6+6q_0)\Omega_{K,0}]z^4+{\mathcal O}(z^5).
\end{aligned}
\end{equation}
To mitigate the convergence problem that the cosmographic expansion encounters at higher redshift (e.g., $z>1$), 
it is useful to recast the $d(z)$ in the form of the \emph{y-redshift}, $y=z/(1+z)$. 
In this manner, the distance in $z\in (0, \infty)$ is mapped into $y\in (0,1)$,  
whereby the cosmographic expansions would be well behaved in the redshift range from the local universe to the ``Big Bang". 
Thus, with the change of variable $z:\to y$, the dimensionless co-moving distance becomes
\begin{equation}
\label{Eq:dcy}
\begin{aligned}
d(y)=y+\frac{1}{2}(1-q_0)y^2+\frac{1}{6}(2-2q_0+3q_0^3-j_0+\Omega_{K,0})y^3 \\
+ \frac{1}{24}[6-6q_0+9q_0^2-15q_0^3-(3-10q_0)j_0 \\
+ s_0+ (6-6q_0)\Omega_{K,0}]y^4+{\mathcal O}(y^5).
\end{aligned}
\end{equation}
With the relation of $ d(z) =H_0d_{\rm L}(z)/c(1+z)$, the luminosity distance in the FLRW universe is given by
\begin{equation}
\label{Eq:dlz}
\begin{aligned}
d_{\rm L}(z)=\frac{c}{H_0}\{z+\frac{1}{2}(1-q_0)z^2 -\frac{1}{6}(1-q_0-3q_0^3  
+j0- \Omega_{K,0})z^3 \\
+\frac{1}{24}[2-2q_0-15q_0^2-15q_0^3+(5+10q_0)j_0 \\
+ s_0 - (2+6q_0)\Omega_{K,0}]z^4+{\mathcal O}(z^5)\}.
\end{aligned}
\end{equation}
In terms of the \emph{y-redshift}, $d_{\rm L}(z)$ is converted to
\begin{equation}
\begin{split}
d_{\rm L}(y)=\frac{c}{H_0}\{y+\frac{1}{2}(3-q_0)y^2 +\frac{1}{6}(11-5q_0+ 3q_0^2-j0+  \Omega_{K,0})y^3 \\
+\frac{1}{24}[50-26q_0 +21 q_0^2-15q_0^3 - (7-10q_0)j_0 \\ + s_0+(10-6q_0)\Omega_{K,0}]y^4+{\mathcal O}(y^5)\}.
\end{split}
\end{equation}
The distance modulus $\mu(z)=5\log [d_{\rm L}(z)/{\rm Mpc}]+25$ can be obtained from the ``logarithmic Hubble relation": $\ln (d_{\rm L}(y)/\, \rm{Mpc})/y=\ln 10/5[\mu(y)-25]-\ln y$, which gives
 \begin{equation}
\begin{split}
\mu (y)=25+\frac{5}{\ln \, 10}\{\ln(c/H_0)+\ln \,y +\frac{1}{2}(3-q_0)y+\frac{1}{24}[17 -
2q_0 \\ +9q_0^2 
- 4(j_0-\Omega_{K,0})]y^2 
+ \frac{1}{24}[11-q_0 +2q_0^2- 10q_0^3 -(1-8q_0)j_0 \\
+s_0+ (16-4q_0)\Omega_{K,0} ]y^3 +{\mathcal O}(y^4)\}.
\end{split}
\end{equation}

\label{lastpage}
\end{document}